# Orowan strengthening with thermal activation

Guangpeng Sun, Mingyu Lei, Sha Liu, Bin Wen*

State Key Laboratory of Metastable Materials Science and Technology, Yanshan University,

Qinhuangdao 066004, China.

**Abstract** Orowan strengthening is a primary way of strengthening metallic materials. The effect of temperature on the Orowan strengthening mechanism is still debatable in the present day. In this study, the effect of temperature on the Orowan strengthening mechanism is systematically investigated, a thermal activation Orowan strengthening mechanism is developed, and the corresponding Orowan stress is deduced. The results indicate that the obstacle scales substantially affect the thermal activation–based dislocation that bypasses the precipitates processes. In small obstacle scales, the thermal activation contribution to Orowan stress cannot be ignored; while in large obstacle scales, the thermal activation contribution can be ignored even at high temperatures. In addition to temperature, the effects of shear modulus, strain rate, and dislocation density are investigated on Orowan stress. This work not only provides new insight into the Orowan strengthening mechanism but also aids in the development of new high-temperature structural materials.

**Keywords:** Orowan strengthening, thermal activation mechanism, Orowan stress, precipitate diameter, interprecipitate spacing,  shear modulus, dislocation density, strain rate

* Corresponding author, E-mail address: wenbin@ysu.edu.cn

# 1. Introduction

The critical resolved shear stress of dislocation slip determines the strength of crystalline materials, which can be enhanced by impeding dislocation movement [1-4]. There are typically four strategies for strengthening, including solution strengthening, dislocation strengthening, grain boundary strengthening, and Orowan strengthening (or precipitates strengthening) [5]. Orowan strengthening is an important strengthening strategy for metallic materials [6-10]. When a dislocation in motion encounters precipitation, its movement is impeded. By increasing the applied stress, the dislocation can bypass or shear the precipitates allowing atoms to slide continually, resulting in Orowan strengthening behavior. The mechanism of Orowan strengthening originates from precipitates; hence, Orowan strengthening can be designed by tailoring precipitates [11-13]. Because it is difficult to tailor the stability of microstructures with high dislocation density and grain boundary density at elevated temperatures, Orowan strengthening becomes a preferred strategy for superalloy strengthening [14]. In addition, the density of moveable dislocations did not drop much during the Orowan strengthening process, indicating that the strengthening can be carried out with less loss of ductility [15]. Consequently, Orowan strengthening is an important superalloy strengthening strategy, and its mechanism has been extensively studied [16-21].

The Orowan strengthening mechanism was first studied in 1948 [1]. Based on the constant line tension model for dislocation slip [22], the stress for a dislocation, bypassing impenetrable precipitates, known as Orowan stress, can be deduced as $Gb/L$, where G is the shear modulus, $b$ is the magnitude of the Burgers vector for dislocation, and L is the interprecipitate spacing. First considered in 1958, the effect of randomly distributed precipitates was stated as $Gb\sqrt{CD}$, where C and D represent the density and diameter of the precipitate [2]. In 1973, based on the effective line tension, Bacon et al. proposed Orowan stress as $(\ln D / \ln L)^{1/2}(Gb/L)(\ln D / 2\pi)$ [16]. In 2001, a further improved model was proposed by Mohles et al. [17, 23], which accounts for the effect of non-regularly distributed precipitates based on an effective precipitate spacing. Overall, the above models represent the Orowan stress to overcome impenetrable obstacles in the absence of thermal activation.

The above models indicated that the Orowan strengthening mechanism is temperature effect independent. However, current research suggests that the effect of temperature on Orowan stress is solely attributable to the influence of temperature on the shear modulus [24, 25], whereas another perspective argues that the influence of thermal activation on Orowan strengthening cannot be neglected [13, 26]. Therefore, the effect of temperature on Orowan strengthening is still a matter of discussion. To resolve this controversy, this study investigates the effect of thermal activation on the Orowan strengthening mechanism and calculates the thermal activation contribution to Orowan stress. According to our calculations, thermal activation contributes more to Orowan stress for short scale obstacles than for long scale obstacles. This work provides new insights into the Orowan strengthening mechanism and will aid in developing high-temperature structural materials.

## 2. Dislocation bowing pathway and Orowan stress for Orowan strengthening with complete mechanical activation

To study the effect of thermal activation on the Orowan strengthening mechanism, it is necessary to generate a thermal activation energy barrier via a dislocation-bowing pathway. This analysis approximates the route to be a dislocation-bowing pathway with total mechanical activation. As depicted schematically in Fig. 1a, for Orowan strengthening with complete mechanical activation, when a dislocation in motion encounters impenetrable precipitates, the motion of dislocation is hindered. By increasing the external shear stress, the dislocation can bypass the impenetrable precipitates. Orowan stress, i.e., the critical external shear stress for a dislocation to bypass impenetrable precipitates, is known as the BKS stress $\tau_{BKS}$ because Bacon, Kocks, and Scattergood studied it [16]. Following the BKS method, the bowing pathway and BKS stress $\tau_{BKS}$ of dislocation are calculated. The following section describes the computational method in greater detail.

As shown in Fig. 1a, the interaction between dislocation branches is very strong near the precipitates, possibly leading to instabilities and pinching-off. Consequently, the previous approximation that the equilibrium bowing loops are simple circular arcs is insufficient [16, 27], and the interaction between dislocation branches must be considered. For an individual bowing loop to reach an equilibrium dislocation configuration, as shown in Fig. 1b, under a given external shear stress $\tau$, each point on the equilibrium bowing loop must fulfill the static stress equilibrium equation (Eq. 1) [16]:

$$\tau + \tau_{self} = 0, \tag{1}$$

where $\tau_{self}$ is the self-stress. According to Brown's work [28], the self-stress can be estimated numerically as the sum of stress over all discrete elements. Thus, the self-stress was determined by the entire bowing configuration and its shear modulus. A numerical method determined the equilibrium dislocation configuration for a given external shear stress. Initially, an original bowing configuration was assumed, and a discrete set of finite points and elements were identified. The self-stress of a set of finite points on the original bowing configuration was then computed. Second, the net-unbalanced stress $\Delta\tau = \tau + \tau_{self}$ was calculated, and it may be used to relax the dislocation to a new bowing configuration by modifying the curvature of the elements following the iteration relationship as follows (Eq. 2):

$$k_{m+1} = k_m [1 - c\Delta\tau / (\tau_{arc})_m], \tag{2}$$

where $k_m$ is the curvature of the elements in this iteration, and $k_{m+1}$ is the curvature of the elements in the next iteration; the constant $c = 0.7$ is to prevent oscillations; $(\tau_{arc})_m$ is the stress contribution from the arc element containing the given points to self-stress in this iteration [16]. The iteration was finally terminated when the net-unbalanced stress was less than 1% of the external shear stress. Using this method, an equilibrium dislocation configuration under a given external shear stress can be obtained [16].

As the external shear stress increases, the equilibrium dislocation configuration continues to bow outward. As shown in Fig. 1c, when the external shear stress reaches a critical stress [16], the dislocation loop becomes unstable and bypasses the precipitates; this critical stress is the BKS stress $\tau_{BKS}$, and this critical dislocation configuration is referred to as the BKS configuration. Due to the absence of temperature effects, the BKS stress is the stress level required to bypass precipitates under complete mechanical activation. Because the dislocation configuration varies with increasing external shear stress, the transition from the initial dislocation configuration to the BKS configuration is known as the bowing pathway. The calculated bowing pathways for edge dislocation, screw dislocation, and mixed (45°) dislocation are shown in Fig. 2a–2c, respectively, and are in good agreement with the literature data over a wide D/L range (Table S1) [16, 27]. In addition, our bowing pathways are extremely compatible with MD-simulated snapshots (Fig. S1). Therefore, the numerical method used in this study to determine the bowing pathway and BKS stress is highly reliable.

## 3. Thermal activation Orowan strengthening mechanism and Orowan stress

Although the BKS stress can be used to quantify the stress level that bypasses impenetrable precipitates under complete mechanical activation, and it is somewhat acceptable at low temperatures, it is questionable that the BKS stress is used to estimate the Orowan stress at high temperatures. In this section, a thermal activation Orowan strengthening mechanism is proposed, and the corresponding thermal-activated Orowan stress is deduced based on the above-calculated dislocation-bowing pathway.

### 3.1 Thermal activation Orowan strengthening mechanism and Orowan stress

As schematically illustrated in Fig. 3a, the Orowan bypassing mechanism is regulated by mechanical and thermal activation. As studied in the above section, when the dislocation loop reaches the BKS configuration, it becomes unstable and bypasses the precipitates. Hence, the precondition for dislocation bypassing precipitates, whether heat or mechanical activation, is that the dislocation configuration reaches the BKS configuration. Given a given external shear stress less than the BKS stress, a dislocation can achieve its equilibrium configuration with complete mechanical activation, but it cannot bypass precipitates. The equilibrium configuration may achieve the BKS configuration through thermal activation. By definition, thermal activation energy is the difference in system energy between the BKS and the equilibrium configurations. The activation energy as a function of the external shear stress $\tau$ can be written as follows (Eq. 3):

$$\Delta\Phi(\tau) = gGb^3[1-(\tau/\tau_{BKS})^p]^q, \qquad (3)$$

where $gGb^3$ is a stress-free energy barrier, and the exponents p and q are empirical parameters.

By combining Eq. 3 with the Orowan equation [29-31] and Arrhenius equation [26, 32, 33], the following constitutive equation can be obtained (Supplementary Part. SIII) [34-36]:

$$gGb^3[1-(\tau/\tau_{BKS})^p]^q = k_B T\ln(\rho b\lambda\omega/\dot{\varepsilon}), \qquad (4)$$

where $k_B$ is the Boltzmann constant, T is the temperature, $\rho$ is the density of mobile dislocations, $\dot{\varepsilon}$ is the plastic strain rate, $\lambda$ is the dislocations mean free path, and $\omega$ is the Debye frequency.

For a given temperature, mobile dislocations density, and plastic strain rate, by simplifying Eq. 4, Orowan stress can be written as the product of (1-$\alpha$) and BKS stress $\tau_{BKS}$, and $\alpha = 1-\left(1-[k_B T\ln(\rho b\lambda\omega/\dot{\varepsilon})/(gGb^3)]^{1/q}\right)^{1/p}$ is defined as the thermal activation factor. Therefore, $\alpha$ can characterize the thermal activation effect on Orowan stress under a given temperature, strain rate, and microstructure condition. Higher the value of α, higher the effect of temperature on Orowan stress and the material more softening.

If the randomness precipitate distribution is considered, thermal-activated Orowan stress can be further written as (Eq. 5):

$$\tau_o = (1-\alpha)\beta\tau_{BKS}, \tag{5}$$

where $\beta = (\ln D/\ln L)^{3/2}$ is the randomness factor of the precipitates [16].

**3.2 Orowan stress from a numerical method**

Theoretically, using Eq. 5, the thermal-activated Orowan stress can be determined. In this equation, the BKS stress can be computed using the method described in Section 2; if the geometric size and distribution of the precipitates are known, the randomness factor of precipitates, β, can be calculated. For the thermal activation factor α, three empirical parameters, g, p, and q, need to be determined. If the activation energy function $\Delta\Phi(\tau)$ is known, all of these empirical parameters may be determined, and the thermal-activated Orowan stress can be calculated using Eq. 5. Additionally, we may easily substitute the left side of Eq. 4 with the activation energy function $\Delta\Phi(\tau)$, and solve Eq. 4 to obtain the thermal-activated Orowan stress. Using the latter method, the thermal-activated Orowan stress was determined.

Based on the dislocation theory [22], the dislocation system potential energy $\Phi(\tau)$ as a function of the external shear stress τ includes four components, namely the elastic energy of dislocation $W_{ela}$, the interaction energy within the bowing configuration $W_{inti}$, the interaction energy from the adjacent bowing configuration $W_{inta}$, and the work done by the external shear stress $W_\tau$. Therefore, $\Phi(\tau)$ can be expressed as follows (Eq. 6):

$$\Phi(\tau) = W_{ela} + W_{inti} + W_{inta} + W_\tau, \tag{6}$$

Based on the bowing pathway obtained in Section 2, $\Phi(\tau)$ can be calculated using a numerical method (supplementary Part. SII). As depicted schematically in Fig. 3b, the dislocation system potential energy along the bowing pathway forms a function with a minimum value (at equilibrium configuration), and a maximum value (at BKS configuration) for a given external shear stress, and the difference between the maximum and minimum values is the activation energy $\Delta\Phi$. As a result, it is possible to obtain the activation energy $\Delta\Phi$ as a function of $\tau$. As shown in Fig. 3c, activation energy decreases with increasing external shear stress. By substituting the left side of Eq. 4 with the function $\Delta\Phi(\tau)$, the thermal-activated Orowan stress $\tau_o$ can be obtained at a given temperature, mobile dislocation density, and plastic strain rate condition.

### 3.3 Orowan stress from the analytical method

Although the thermal-activated Orowan stress $\tau_o$ can be obtained by numerically calculating the activation energy function $\Delta\Phi(\tau)$, the calculation technique is relatively difficult and requires a significant amount of processing resources. Next, we attempted to solve Eq. 4 using empirical parameters directly. According to Brown's work [16, 28], the dimensionless parameters p and q can be set as constants, that is, p = 1, and q = 1.5 [37-39], and the dimensionless parameter g is only related to the D and L [5]. $g = \ln(D)\ln(L)\ln[DL/(D+L)](1-\nu\cos^2\gamma)/(2\pi-2\pi\nu)$ was selected in this work, where $\nu$ is Poisson's ratio and $\gamma$ is the angle between the dislocation line and the Burgers vector. Using these empirical parameters, the activation energy as a function of external shear stress can be estimated using Eq. 3, and this result is in good agreement with those numerically derived in Section 3.2 throughout a wide D/L ratio range (see Fig. S2). Consequently, the thermal activation factor $\alpha$ can be obtained directly by substituting these empirical parameters into its expression; furthermore, the thermal-activated Orowan stress under thermal-mechanical activation and random distribution mechanism can be obtained directly via Eq. 5.

## 4. Verification of thermal-activated Orowan stress model

To validate the rationale and reliability of this thermal-activated Orowan stress model, it is used in several nano-oxide dispersion-strengthened steels and nickel-based superalloys, which are typical Orowan strengthening materials with impenetrable precipitates [13].

### 4.1 Comparison of computed thermal-activated Orowan stress with experimental results

Using the method described in Section 3.2 and the parameters listed in Table 1, the thermal-activated Orowan stresses of several alloys were calculated. Adding forest strengthening and solid solution strengthening contribution to flow stress yields their flow stresses, displayed in Fig. 4. The calculated flow stresses are remarkably consistent with the experimental ones [13, 40-45]. As shown in Fig. 4a and 4c, the Orowan strengthening contribution to flow stress (thermal-activated Orowan stress multiplied by the Taylor factor 3.1 [46]) continuously decreases with increasing temperature. This occurs for two reasons: the temperature decreases shear modulus and increases the thermal activation effect, decreasing Orowan stress. As shown in Fig. 4b and 4d, the calculated flow stress for a wide strain rate range ($10^{-2}$–10 s$^{-1}$) or a wide obstacle scale range is also in good agreement with experimental data, indicating that our thermal-activated Orowan stress model can accurately represent the effect of temperature, obstacle scale, and strain rate on Orowan stress. This thermal-activated Orowan stress model may accurately estimate the Orowan stress by reasonably introducing the thermal activation effect.

### 4.2. Comparison between numerical and analytical methods

To verify the reliability of the analytical method, the thermal-activated Orowan stress is calculated with different precipitate diameters and interprecipitate spacings using the numerical and analytical methods described in Sections 3.2 and 3.3. For this calculation, the parameters $G = 75[1-0.5(T-300)/1700]$ GPa [47, 48], $b$ = 0.25 nm, $\nu$ = 0.3, $\rho = 10^{13}$ m$^{-2}$, $\dot{\varepsilon} = 10^{-5}$s$^{-1}$, $\omega$ = 13.5 THz were selected. As shown in Fig. S2 and S3, the thermal-activated Orowan stresses by the analytical calculation method correspond well with the numerical method, indicating the reliability of the analytical method, and the empirical parameters chosen for this study are suitable. This analytical method offers a straightforward method for studying the thermal-activated Orowan stress.

## 5. Thermal-activated Orowan stress and thermal activation factor

The thermal-activated Orowan stress and the thermal activation factor are related to the precipitate diameter, interprecipitate spacing, strain rate, mobile dislocation density, and shear modulus, among other variables. The effect of these parameters on Orowan stress and thermal activation factors is systematically studied to elucidate this correlation.

### 5.1 Effect of precipitate geometrical properties on thermal-activated Orowan stress and thermal activation factor

To investigate the effect of precipitate geometrical properties on $\tau_o$ and $\alpha$, the activation energy as a function of external shear stress is first studied using the numerical method for different D and L. Here, G = 75 GPa, $b$ = 0.25 nm, $\nu$ = 0.3 are chosen to simulate FCC copper alloy metal. The calculated activation energy for an edge dislocation is displayed in Fig. 5a and 5b. The results for screw and mixed (45°) dislocation are shown in Fig. S4 and S5, respectively. In general activation energy decreases with increasing external shear stress, and the decrease is steeper for a large obstacle scale (D+L) instance; with high steeper, the activation area becomes wider (Fig. 5c and 5d). The interaction energy ($W_{inta}$, $W_{inti}$) contributes less to $\Phi$ in the case of large barrier scales, as determined by evaluating the energy component of potential energy of the dislocation system $\Phi$ at different D and L.

To examine the effect of precipitate geometrical properties on thermal-activated Orowan stress, the thermal-activated Orowan stresses as a function of temperature at L = 50$b$ and 300$b$ for an edge dislocation are calculated and displayed in Fig. 6a and 6b as compared with BKS stress [16], and the results for screw and mixed (45°) dislocation are shown in Fig. S6 and S7. The shear modulus changes with temperature T as $G = 75[1-0.5(T-300)/1700]$ GPa, although the other parameters remain constant, i.e., $\dot{\varepsilon} = 10^{-2}$ s$^{-1}$, $\rho = 10^{12}$ m$^{-2}$, $\lambda = 10b$, and $\omega = 10^{13}$ s$^{-1}$. As observed, both the thermal-activated Orowan stress and the BKS stress decrease with increasing temperature, indicating that temperature-dependent shear modules affect the Orowan stress. However, the calculated thermal-activated Orowan stress is less than the BKS stress at elevated temperatures, indicating that thermal activation affects Orowan stress further. In the instances of L = 50$b$ and 300$b$, the thermal-activated Orowan stress decreases with increasing D; however, under the same D condition, the thermal-activated Orowan stress for L = 50$b$ is greater than that for L = 300$b$. Meanwhile, a large D will increase the $\tau_{BKS}$, while a large L will decrease the $\tau_{BKS}$.

Fig. 6c and 6d display the thermal activation factor α as a function of temperature for edge dislocation at L = 50$b$ and 300$b$. The α for screw and mixed (45°) dislocation is shown in Figs. S6 and S7. Under given D and L, the α increases with increasing temperature, indicating that the thermal activation contribution increases with increasing temperature. In addition, at a constant temperature, greater values of D and L result in a slower increase in the α, showing that the thermal activation effect contributes less in the case of greater D and L. The above analysis demonstrates that large values of D and L weaken the thermal activation effect, resulting in a small α. To prevent softening by high temperature, a small α value is required. To improve the strengthening effect, a large $\tau_{BKS}$ value and a small α value are required indicating that a large precipitate diameter and a small interprecipitate spacing are required; this result is consistent with the observation of the superalloy microstructure [2, 16].

**5.2 Effect of precipitate-volume fraction on thermal-activated Orowan stress and thermal activation factor**

The precipitate-volume fraction is a technique parameter utilized more frequently in the real Orowan strengthening strategy. This section discusses the effect of the precipitate-volume fraction on the thermal-activated Orowan stress. In this calculation, shear modulus G = 75[1-0.5(T-300)/1700] GPa, $\dot{\varepsilon}$ = 10$^{-2}$ s$^{-1}$, ρ = 10$^{12}$ m$^{-2}$, λ = 10$b$, and ω = 10$^{13}$ THz were chosen. The calculated thermal-activated Orowan stress for edge dislocation is shown in Fig. 7, and the results for screw and mixed (45°) dislocation are shown in Fig. S8 and S9. The interprecipitate spacing L can be calculated from the precipitate diameter D and precipitate-volume fraction $c$ using the formula L = ($\sqrt[3]{\pi/(6c)}$-1)D [5]. As shown in Fig. 7a, L increases with increasing D at a constant precipitate-volume fraction. Fig. 7b displays that the thermal-activated Orowan stress increases with decreasing D when $c$ remains constant, which is highly consistent with the experimental observation [2], and that the thermal-activated Orowan stress increases with increasing precipitate-volume fraction at a constant D and temperature. By evaluating the L at a constant precipitate-volume fraction, it was found that the L decreases as D decreases and that it dominates the increase in the thermal-activated Orowan stress.

The thermal activation factor, α as a function of temperature, precipitate-volume fraction, and precipitate diameter is displayed in Fig. 7c, and the results for screw and mixed (45°) dislocation are displayed in Figs. S8b and S9b, respectively. The thermal activation factor increases with increasing precipitate-volume fraction $c$ at constant temperature and D. At a constant precipitate-volume fraction, the thermal activation factor increases with decreasing precipitate diameter. Similar to Section 5.1, the increase of α is due to the small D and L, which contribute more to the thermal activation effect. At constant $c$, although small D may induce the increase of α, the subsequent small L will lead to an increase in $\tau_{BKS}$. Therefore, in most experimental observations [5], the precipitation strengthening effect increases as D decreases.

**5.3 Effect of shear modulus on thermal-activated Orowan stress and thermal activation factor**

According to dislocation theory, there is a strong correlation between material strength and shear modulus [34]. The thermal-activated Orowan stress and thermal activation factor were then calculated as a function of shear modulus. In this calculation, the parameters $b$ = 0.25 nm, L = 50$b$, D = 5$b$, $\nu$ = 0.3, $\dot{\varepsilon}$ = $10^{-2}$ s$^{-1}$, $\rho$ = $10^{12}$ m$^{-2}$, $\lambda$ = 10$b$, and $\omega$ = $10^{13}$ THz were chosen. Fig. 8a displays the temperature-dependent thermal-activated Orowan stress under different shear modulus cases for an edge dislocation, and the results for screw and mixed (45°) dislocation are shown in Fig. S10. As observed, the thermal-activated Orowan stress decreases with increasing temperature for all shear modulus cases. In contrast to the low shear modulus, the high shear modulus consistently exhibits a large thermal-activated Orowan stress over the whole temperature range. By evaluating their static stress equilibrium equation (Eq. 1), it has been determined that the self-stress increases linearly with an increase in the shear modulus [16], increasing the thermal-activated Orowan stress.

To measure the effect of shear modulus on the thermal activation factor, $\alpha$, Fig. 8b displays the thermal activation factor as a function of temperature for various shear modulus values. The low shear modulus typically exhibits a high thermal activation factor than a high sheer modulus. The activation energy decreases linearly with decreasing shear modulus, as determined by assessing the activation energy at constant D and L (Eq. 3 or Eq. S6). Consequently, at a constant temperature, a low shear modulus can result in a large $\alpha$. Based on the above analysis, if someone wishes to develop an anti softening, high-strength material at high temperatures, a high shear modulus is essential, which not only increases the $\tau_{BKS}$ but also decreases $\alpha$.

## 5.4 Effect of mobile dislocation density and strain rate on thermal-activated Orowan stress and thermal activation factor

In addition to the parameters mentioned above, mobile dislocation density and strain rate can demonstrate an impact on the thermal-activated Orowan stress. To study the effects of these parameters, thermal-activated Orowan stress under different mobile dislocation densities and strain rate conditions for an edge dislocation is calculated and shown in Fig. 9. The results for screw and mixed (45°) dislocation are shown in Fig. S11 and S12, respectively. In this calculation, the parameters are $b$ = 0.25 nm, L = 50$b$, D = 5$b$, $\nu$ = 0.3, $\lambda$ = 10$b$, $\omega$ = $10^{13}$ THz, G = 75[1-0.5(T-300)/1700] GPa. As observed, the thermal-activated Orowan stress decreases with increasing temperature. Moreover, at a given temperature, the thermal-activated Orowan stress decreases with increasing mobile dislocation density due to the high mobile dislocation to a large thermal activation factor $\alpha$ (Eq. 4), as shown in Fig. 9a and 9b. In addition, the increasing strain rate decreases the thermal activation factor $\alpha$ (Eq. 4). Thus, the thermal-activated Orowan stress increase with increasing strain rate, as shown in Fig. 9c and 9d. These patterns are comparable to the results of the other simulation [13, 49, 50].

## 6. Thermal activation contribution to thermal-activated Orowan stress under different obstacle scales and temperature

Based on the above-proposed thermal-activated Orowan strengthening mechanism and the analysis, the temperature effect on the Orowan strengthening mechanism and Orowan stress was systematically investigated. Numerous factors were found to affect Orowan stress. In particular, obstacle scale and temperature have an important effect on the thermal-activated Orowan stress. Fig. 10 is a schematic depiction of a dislocation bypassing mechanism map in terms of the obstacle scale and temperature that reflects the effect of these elements on Orowan stress. As observed, the contribution of thermal activation can be altered by adjusting the obstacle scales and temperature. For small obstacle scales, the thermal activation strongly influences the dislocation bypassing process, resulting in a fast decrease in thermal-activated Orowan stress with increasing temperature or a fast increase in thermal activation factor;

whereas for large obstacle scales, the thermal activation contribution can be neglected even at high temperature, resulting in a slow decrease in the thermal-activated Orowan stress.

Following Eq. 5, the thermal-activated Orowan stress $\tau_o$ consists of two components: the mechanical activation component $\beta\tau_{BKS}$ [16] and the thermal activation component. The temperature has no effect on the mechanical activation component $\beta\tau_{BKS}$, and its effect is wholly concentrated on the thermal activation factor $\alpha$; therefore, the effect of temperature on Orowan stress can be determined by analyzing the thermal activation factor, $\alpha$. Consequently, while designing superalloys, selecting an obstacle scale must consider the effects of thermal and mechanical activation. For a material with a shear modulus of 100 GPa, if D is greater than ~100 nm and L is greater than ~100 nm, their thermal activation factor at a temperature of 1500 K is less than 0.1; this indicates that the thermal activation effect can be ignored; otherwise, it cannot be ignored.

In addition to the obstacle scale, the shear modulus considerably impacts thermal activation-based bypassing processes. When the shear modulus exceeds ~200 GPa, the mechanical activation component controls the bypassing of dislocation. However, as the shear modulus decreases below 25 GPa, the thermal activation component even dominates the reduction of the thermal-activated Orowan stress.

## 7. Conclusions

In conclusion, an Orowan strengthening mechanism with thermal activation is proposed, and a thermal-activated Orowan stress expression is developed in this work. Comparing the calculated results with experimental data in numerous nano-oxide dispersion-strengthened alloys verifies the rationality and reliability of this thermal-activated Orowan stress model. The results indicate that the obstacle scales substantially affect the thermal activation-based dislocation that bypasses the precipitates processes. Even at high temperatures, the thermal activation contribution to Orowan stress cannot be neglected at small obstacle scales, although it may be ignored at large obstacle scales. Our results also indicate that the temperature effect on Orowan stress increases with increasing temperature and decreasing shear modulus, indicating a feasible tailoring strategy to optimize the mechanical properties of engineered materials. This study contributes to clarify the physics underlying the Orowan strengthening mechanism and guides the design of the superalloy.

**Acknowledgment:** This work was supported by the National Natural Science Foundation of China (Grant Nos. 51925105, 51771165) and the National Key R&D Program of China (YS2018YFA070119).

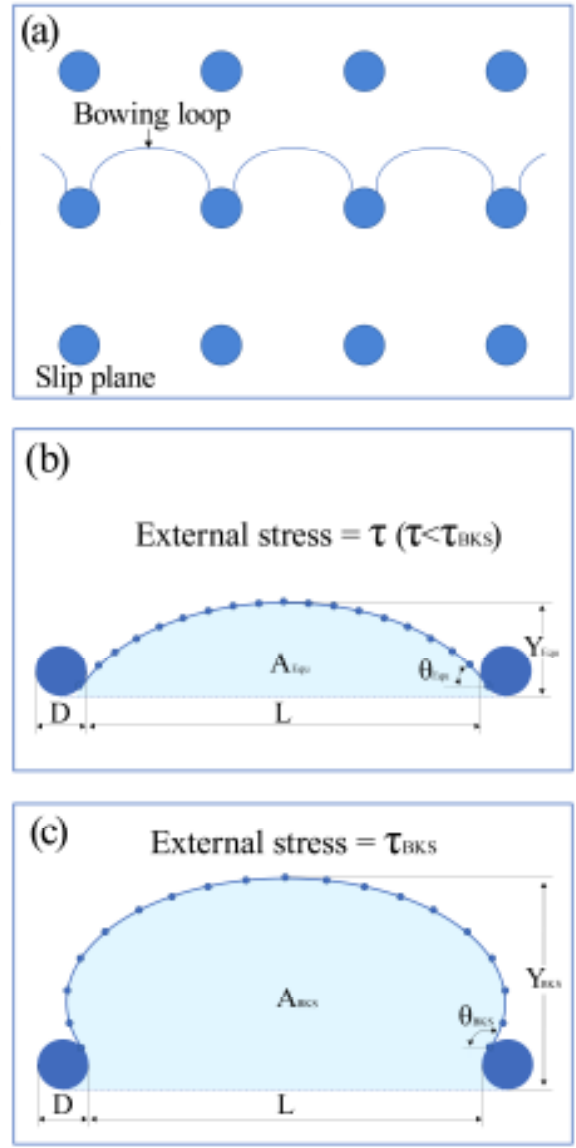

**Fig. 1. Schematic diagram of a computational method for a dislocation configuration under external shear stress.** (a) Schematic illustrating that dislocation is hindered by some impenetrable spherical precipitates. A line represents the dislocation in the slip plane and spheres represent the impenetrable precipitates, the precipitate diameter is D, and the interprecipitate spacing is L. (b) Equilibrium dislocation configuration under external shear stress τ; here, the external shear stress τ is less than BKS stress $\tau_{BKS}$. BKS stress $\tau_{BKS}$ is the critical shear stress that a dislocation bypasses the precipitates through complete mechanical activation. The line segment represents the discrete elements used for finite element calculation. $A_{Equ}$, $\theta_{Equ}$, and $Y_{Equ}$ is swept area, tangent angle, and height of the bowing loop for equilibrium configuration, respectively. (c) The dislocation configuration under BKS stress $\tau_{BKS}$, is called BKS dislocation configuration. $A_{BKS}$, $\theta_{BKS}$, and $Y_{BKS}$ are swept area, tangent angle, and height of the bowing loop for BKS dislocation configuration, respectively.

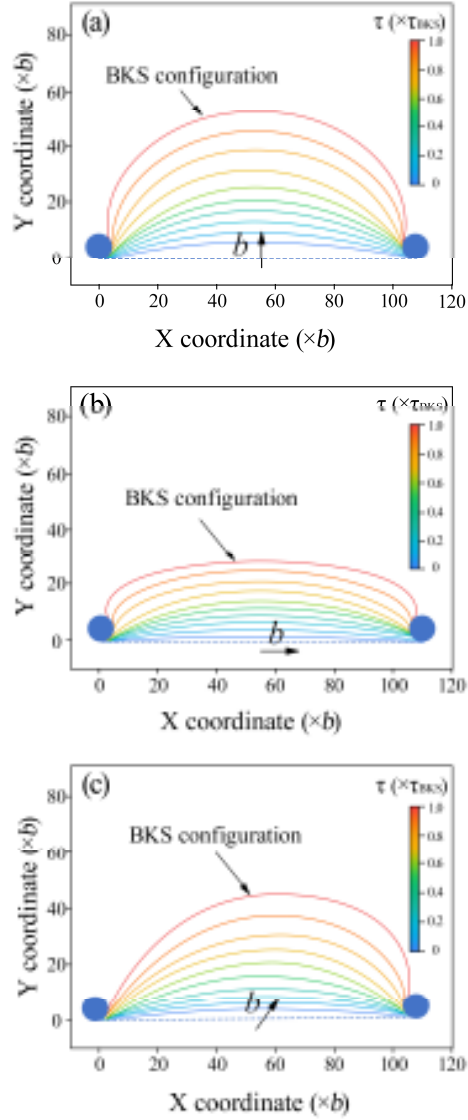

**Fig. 2. Equilibrium and BKS (critical) dislocation configurations from the numerical calculation method.** (a) The equilibrium dislocation configurations for edge dislocation at increasing external shear stress $\tau$ and the corresponding BKS dislocation configuration. The dashed line denotes the original straight dislocation, the outermost curve represents the BKS dislocation configuration, and others represent the equilibrium dislocation configuration. The increment of external shear stress between configurations is $0.1\tau_{BKS}$. The external shear stress is always parallel to the Burger vector. The precipitate diameter is D = 10$b$, the interprecipitate spacing is L = 100$b$, and $b$ is the magnitude of the Burger vector. (b) The equilibrium dislocation configurations for a screw dislocation at increasing external shear stress $\tau$ and the corresponding BKS dislocation configuration. (c) The equilibrium dislocation configurations for mixed (45°) dislocation at increasing external stress $\tau$ and the corresponding BKS dislocation configuration.

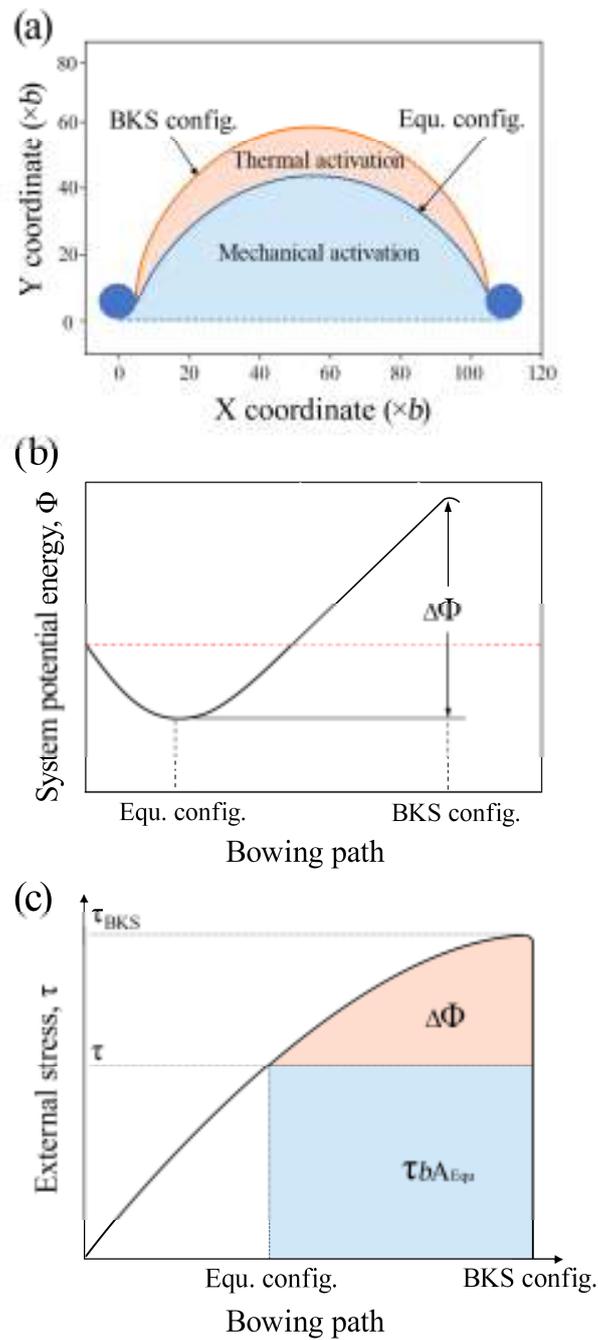

**Fig. 3. Thermal activation mechanism for Orowan strengthening.** (a) Schematic illustration of the Orowan bypassing mechanism that is controlled by the mechanical and thermal activations. (b) System potential energy as a function of the dislocation bowing pathway under a given external shear stress. $\Delta\Phi$ is the activation energy for dislocation bypassing the precipitates, which is given by the difference between the local maximum and the local minimum values of a potential energy function. (c) Relationship between the thermal and mechanical activation energy under external shear stress $\tau$.

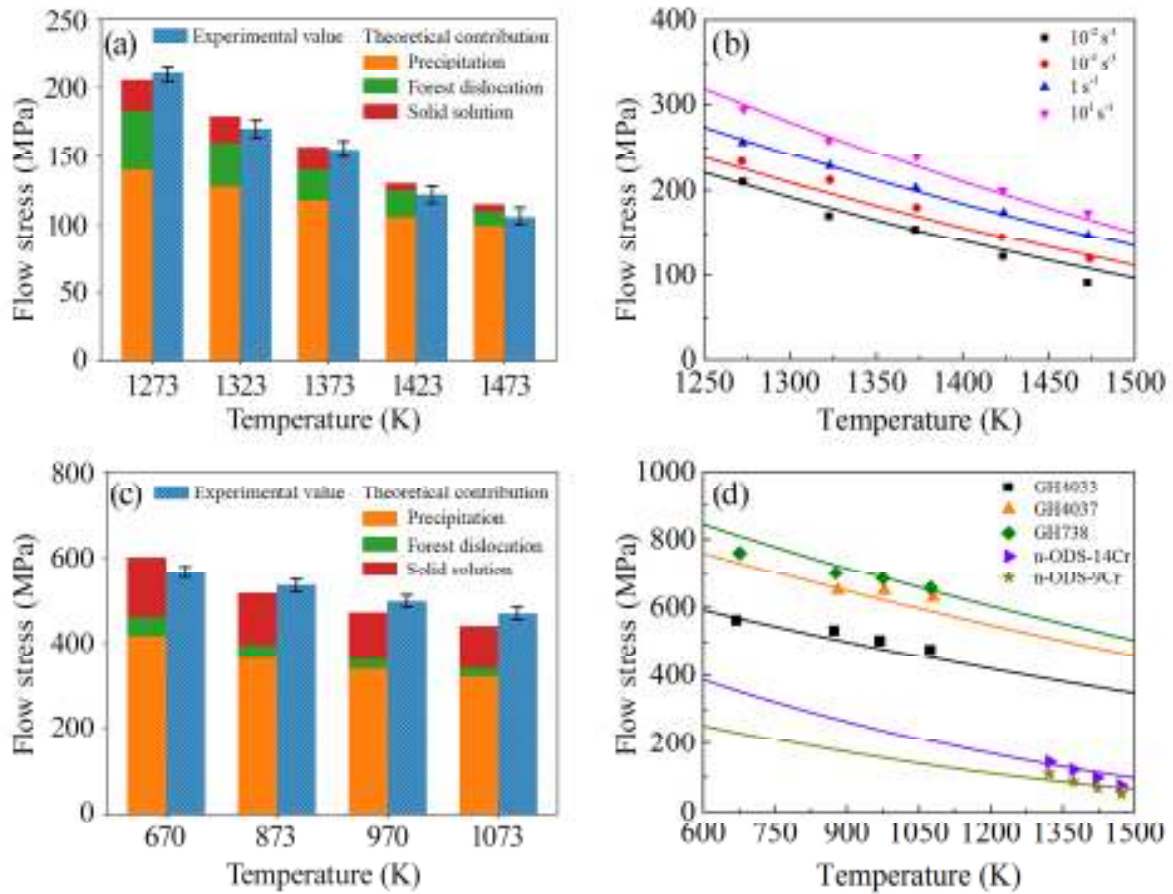

**Fig. 4. Comparison between the reported experimental flow stress and this work's theoretical predicted value.** (a) The contribution of Orowan stress to flow stress in n-ODS-18Cr steel with strain rate $10^{-2}$ s$^{-1}$. From the literature [13], the flow stress corresponds to the stress of the material at a true strain of 0.28. (b) Our calculated flow stress for n-ODS-18Cr steel with variable strain compared to experimental data [13]. (c) The contribution of Orowan stress to flow stress in GH4033 superalloy with strain rate $10^{-5}$ s$^{-1}$. (d) Our calculated flow stress for n-ODS-9Cr steel [43, 44], n-ODS-14Cr steel [45], and nickel-based superalloys GH4033, GH4037, and GH738 in comparison with experimental data [40-42, 51].

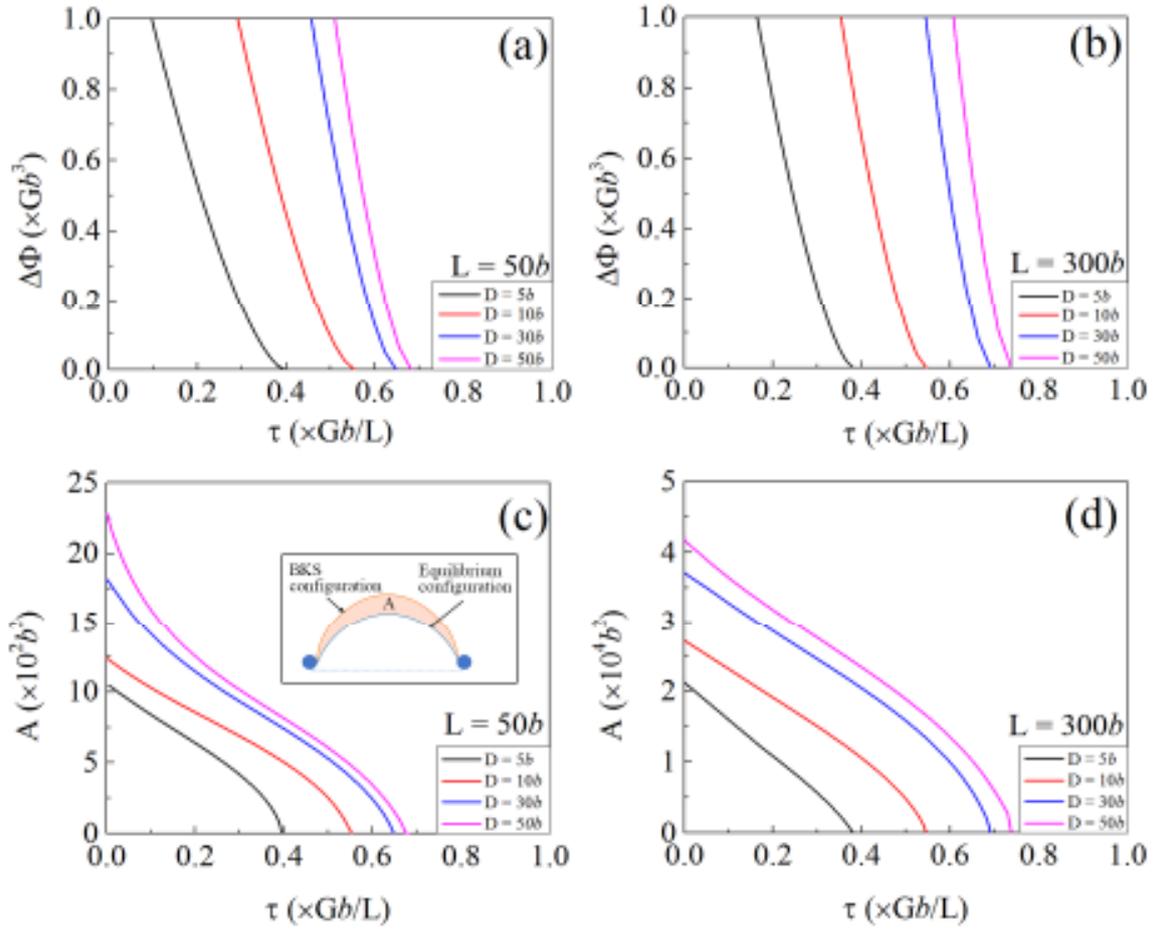

**Fig. 5. The effect of obstacle scale on activation energy and activation area.** (a) Effect of the precipitate diameter on the activation energy for small obstacle scale (L = 50*b*) case. (b) Effect of the precipitate diameter on the activation energy for large obstacle scale (L = 300*b*) case. (c) Effect of the precipitate diameter on the activation area for small obstacle scale (L = 50*b*) case. Insert figure shows the activation area between the BKS and the equilibrium configurations. (d) Effect of the precipitate diameter on the activation area for large obstacle scale (L = 300*b*) case.

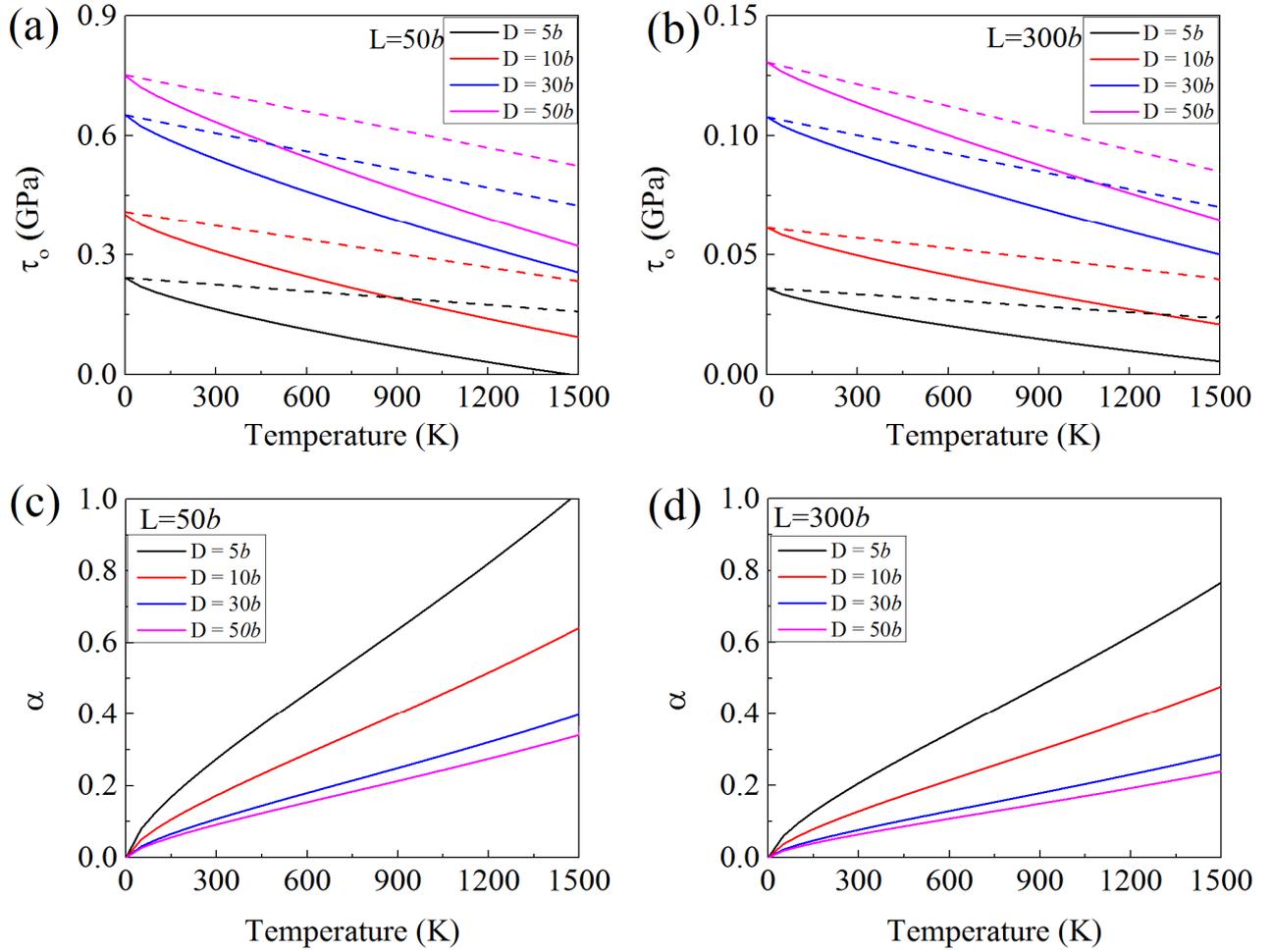

**Fig. 6. The effect of obstacle scale on thermal-activated Orowan stress $\tau_o$ and thermal activation factor $\alpha$.** (a) Effect of precipitate diameter on $\tau_o$ a small obstacle scale ($L = 50b$) case. In this calculation, the mobile dislocation density is $\rho = 10^{12}$ m$^{-2}$, the strain rate is $\dot{\varepsilon} = 10^{-2}$ s$^{-1}$, the shear module as a function of temperature T is $G = 75[1-0.5(T-300)/1700]$ GPa, dislocation mean free path is $\lambda = 10b$, and Debye frequency is $\omega = 10^{13}$ THz. The solid curves are thermal-activated Orowan stress $\tau_o$, calculated by our model, while the dashed lines are Orowan stress without thermal activation, $\beta\tau_{BKS}$ [16]. (b) Effect of the precipitate diameter on $\tau_o$ for large obstacle scale ($L = 300b$) case. (c) Effect of precipitate diameter on $\alpha$ for small obstacle scale ($L = 50b$) case. (d) Effect of the precipitate diameter on $\alpha$ for large obstacle scale ($L = 300b$) case.

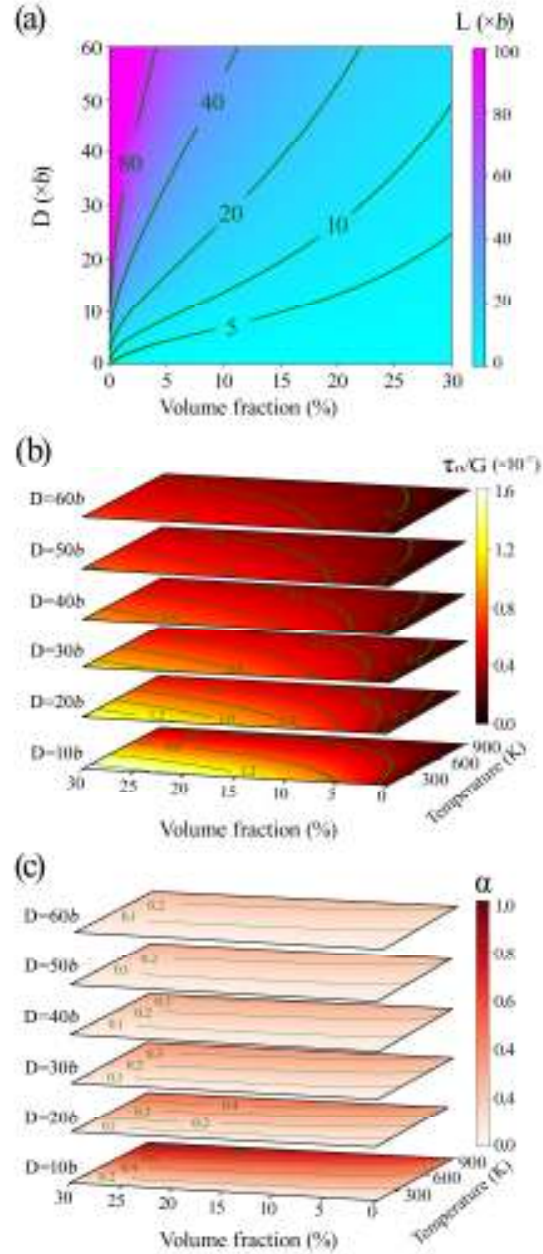

**Fig. 7. Thermal-activated Orowan stress $\tau_o$ and thermal activation factor $\alpha$ under variable precipitate-volume fraction $c$.** (a) Interprecipitate spacing L as a function of precipitate diameter D and precipitate-volume fraction $c$, and this follows the relationship $L = (\sqrt[3]{\pi/(6c)}-1)D$ [5]. (b) Normalized thermal-activated Orowan stress $\tau_o/G$ as a function of precipitate diameter, precipitate-volume fraction, and temperature. In this calculation, the mobile dislocation density is $\rho = 10^{12}$ m$^{-2}$, the strain rate is $\dot{\varepsilon} = 10^{-2}$ s$^{-1}$, shear modulus as a function of temperature T is $G = 75[1-0.5(T-300)/1700]$ GPa, dislocation mean free path is $\lambda = 10b$, and Debye frequency is $\omega = 10^{13}$ THz. (c) $\alpha$ as a function of precipitate diameter, precipitate-volume fraction, and temperature.

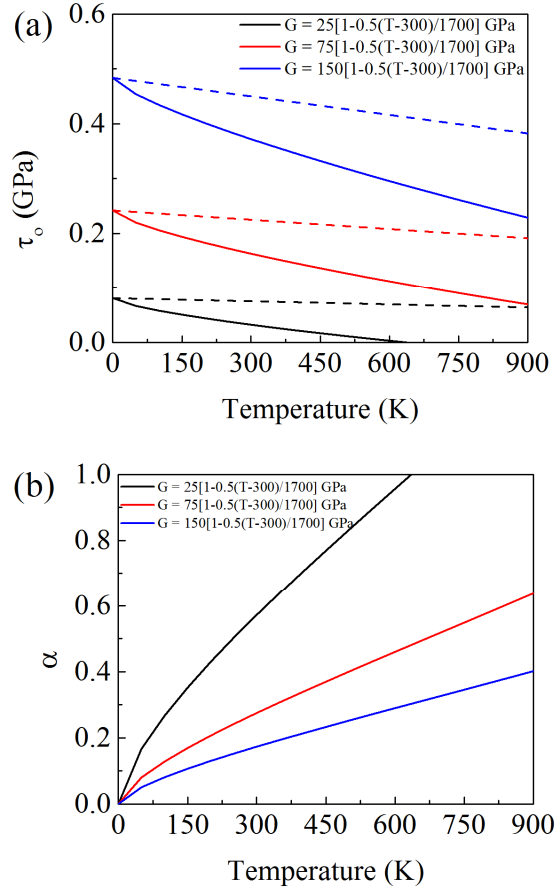

**Fig. 8. Thermal-activated Orowan stress $\tau_o$ and thermal activation factor $\alpha$ under variable shear modulus.** (a) Variations in $\tau_o$ temperature under different shear modulus cases. In this calculation, the mobile dislocation density is $\rho = 10^{12}$ m$^{-2}$, the strain rate is $\dot{\varepsilon} = 10^{-2}$ s$^{-1}$, the precipitate diameter is D = 5$b$, interprecipitate spacing is L = 50$b$, dislocation mean free path is $\lambda$ = 10$b$, and Debye frequency is $\omega$ = $10^{13}$ THz. The solid curves are thermal-activated Orowan stress $\tau_o$ calculated by our model, while the dashed lines are Orowan stress without thermal activation, $\beta\tau_{BKS}$ [16]. (b) Variations in $\alpha$ with temperature under different shear modulus cases.

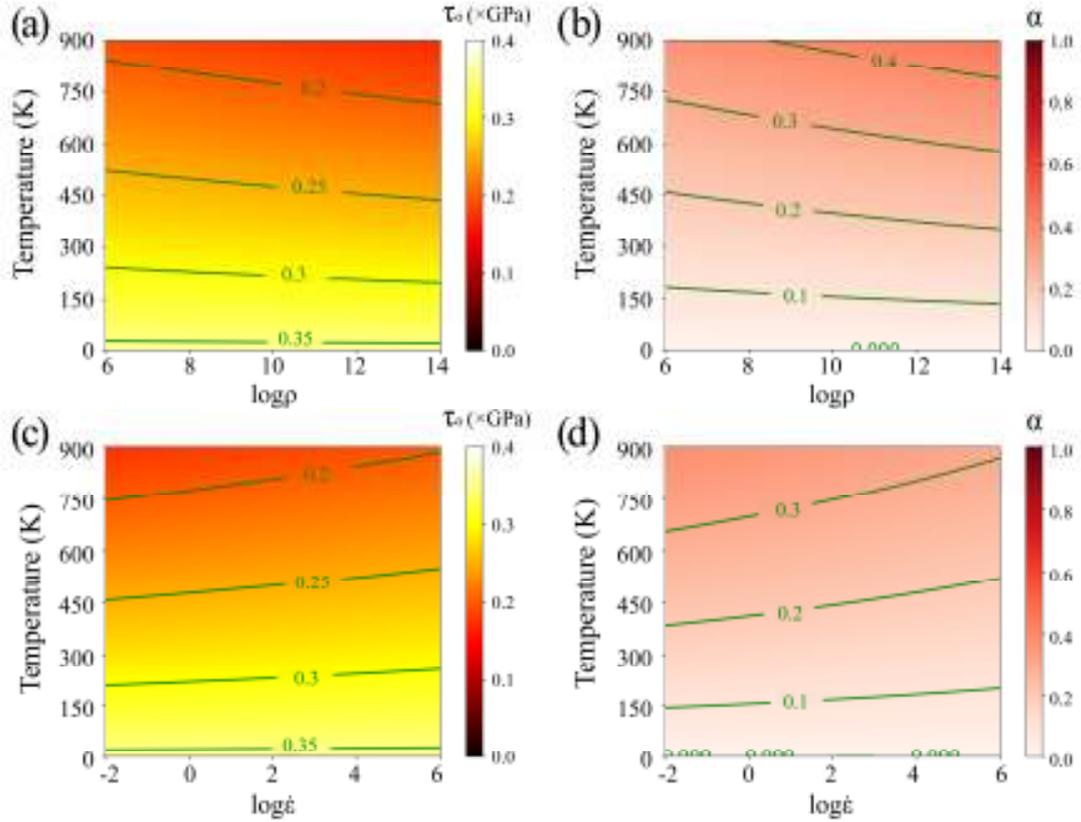

**Fig. 9. Thermal-activated Orowan stress $\tau_o$ and thermal activation factor α as a function of mobile dislocation density and strain rate.** (a) $\tau_o$ as a function of dislocation density and temperature. In this calculation, the mobile dislocation density varied from $10^6$ to $10^{14}$ m$^{-2}$, the temperature varies from 0 to 900 K, and stain rate is $\dot{\varepsilon} = 10^{-2}$ s$^{-1}$, precipitate diameter is D = 10$b$, interprecipitate spacing is L = 50$b$, shear modulus as a function of temperature T is G = 75[1-0.5(T-300)/1700] GPa, dislocation mean free path is λ = 10$b$, Debye frequency is ω = $10^{13}$ THz. (b) α as a function of mobile dislocation density and temperature. (c) $\tau_o$ as a function of strain rate and temperature. In this calculation, the strain rate varies from $10^{-2}$ to $10^6$ s$^{-1}$, and the temperature varies from 0 to 900 K, mobile dislocation density is ρ = $10^{12}$ m$^{-2}$, precipitate diameter is D = 10$b$, interprecipitate spacing is L = 50$b$, the shear modulus as a function of temperature T is G = 75[1-0.5(T-300)/1700] GPa, dislocation mean free path is λ = 10$b$, and Debye frequency is ω = $10^{13}$ THz. (d) α as a function of strain rate and temperature.

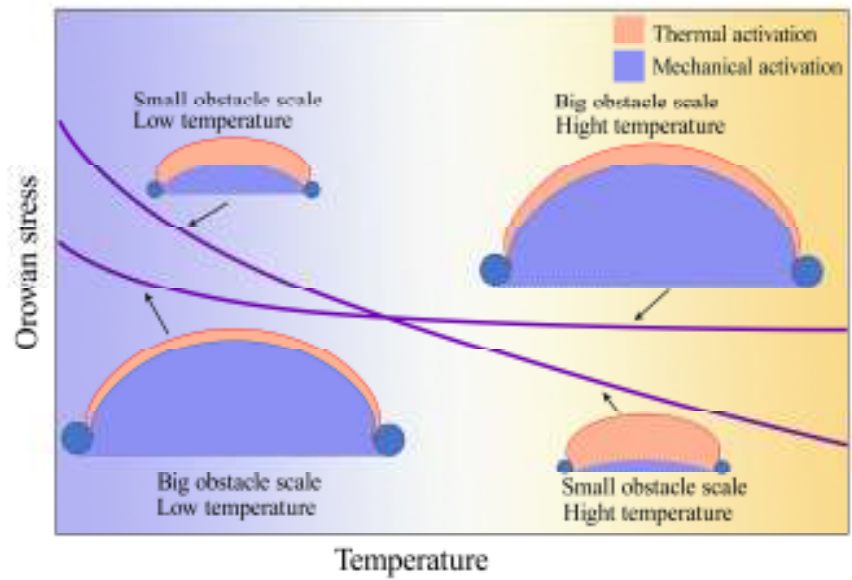

**Fig. 10. Schematic of thermal and mechanical activation contributions to Orowan stress as a function of obstacle scale and temperature.** For a small obstacle scale case, the diameter of precipitate D and the interprecipitate spacing L is small; for a large obstacle scale case, both the diameter of precipitate D and the interprecipitate spacing L are large. With increasing temperature, the thermal activation contribution to Orowan stress increased, and the ratio of thermal activation contribution to Orowan stress is larger for small obstacle scale cases than the large obstacle scale cases.

**Table 1. Parameters used to calculate thermal-activated Orowan stress and forest strengthening and solid solution strengthening contribution to flow stress for different alloys.** D is the precipitate diameter with a unit in nm, L is interprecipitate spacing with a unit in nm, the shear modulus G changes with temperature T as $G_{rt}[1-\delta(T-300)/T_m]$ with a unit in GPa, $b$ is the magnitude of Burgers vector for dislocation with a unit in nm, $\rho$ is mobile dislocations density with a unit in m$^{-2}$, $\dot{\varepsilon}$ is stain rate with a unit in s$^{-1}$, $\lambda$ dislocation mean free path, $\omega$ is Debye frequency with a unit in THz, $\nu$ is Poisson's ratio, $\sigma_{dis}$ is forest strengthening contribution to flow stress with a unit in MPa, and $\sigma_{ss}$ is solid solution strengthening contribution to flow stress with a unit in MPa. The values of $\sigma_{dis}$ and $\sigma_{ss}$ for n-ODS steels are linearly reduced in the temperature range of 1273–1473 K, and the value for nickel-based superalloy is linearly reduced in the temperature range of 600 K–1500 K.

| Alloys / Parameters | n-ODS-18Cr [13, 52, 53] | n-ODS-14Cr [45, 52, 53] | n-ODS-9Cr [44, 52, 53] | GH4033 [40] | GH4037 [40] | GH738 [40] |
|---|---|---|---|---|---|---|
| D | 2 | 7.5 | 10 | 20 | 100 | 65 |
| L | 35 | 40 | 50 | 77 | 80 | 30 |
| $G_{rt}$ | 89.5 | 80 | 82 | 84.6 | 86.5 | 86.2 |
| $\delta$ | 0.84 | 0.84 | 0.84 | 0.54 | 0.54 | 0.54 |
| $T_m$ | 1787 | 1787 | 1787 | 1600 | 1619 | 1633 |
| $b$ | 0.25 | 0.25 | 0.25 | 0.25 | 0.25 | 0.25 |
| $\rho$ | $10^{13}$ | $10^{13}$ | $10^{13}$ | $10^{13}$ | $10^{13}$ | $10^{13}$ |
| $\dot{\varepsilon}$ | $10^{-2} \sim 10^{1}$ | $10^{-2}$ | $10^{-2}$ | $2.8 \times 10^{-5}$ | $10^{-5}$ | $10^{-5}$ |
| $\lambda$ | $10b$ | $10b$ | $10b$ | $10b$ | $10b$ | $10b$ |
| $\omega$ | 12.1 | 12.1 | 12.1 | 12.1 | 12.1 | 12.1 |
| $\nu$ | 0.3 | 0.3 | 0.3 | 0.3 | 0.3 | 0.3 |
| $\sigma_{dis}$ | 60-20 | 60-20 | 60-20 | 141-111 | 144-113 | 143-111 |
| $\sigma_{ss}$ | 27-17 | 27-17 | 27-17 | 75-25 | 77-27 | 76-26 |

# Supplementary Materials for

## Orowan strengthening with thermal activation

Guangpeng Sun, Mingyu Lei, Sha Liu, Bin Wen*

*Corresponding author. Email: wenbin@ysu.edu.cn

**This PDF file includes:**

Part. SI. Simulation Details of Molecular Dynamics

Part. SII. The Derivation of Dislocation System Potential Energy

Part. SIII. The Derivation of Thermal Activation Energy

Supplementary Figures S1 to S12

Supplementary Table S1

**Part. SI. Simulation Details of Molecular Dynamics**

First, a crystal cell was built by redefining a fcc Cu cell [1,1,0], [-1,1,1], and [1,-1,2] directions as X-, Y-, and Z-axes, respectively. And a series of supercell structures that contained ~1.2-2.4 million atoms was constructed by producing the new defined crystal cell 100 × 34 × 18, 100 × 34 × 23, 100 × 34 × 29. In these supercell structures, an edge dislocation along [1,-1,2] orientation with *b*=1/2[110] was introduced by using Atomsk [1]. Second, the MD simulations were conducted for these structures by using a LAMMPS code [2]. The interatomic potential were described by an EMA potential for Cu [3], which has been used to analyse dislocation loops [4]. Periodic boundary conditions were prescribed along the X- and Z-axes, which imitated the dislocation encountering a row of equidistantly precipitates. Margin regions with a length of 2 nm along the Y-axes were set above the bottom and below the top of these supercell structures, which enabled setting a free boundary condition along this direction. In addition, a rigid spherical region simulating impenetrable precipitate with a diameter range of 2.56 ~ 12.8 nm was set on glide plane in above supercell structures. A shear strain along Z-axes applied to the supercell structure in a strain rate of $8.0 \times 10^6$ s$^{-1}$. In these calculations, the NVE ensemble was used, and the time step is set to 10 fs. Finally, the dislocation configurations were visualized and analysed by using DXA method in the OVITO package [5].

**Part. SII. The Derivation of Dislocation System Potential Energy**

Based on dislocation theory [6], the dislocation system potential energy of an equilibrium dislocation configuration was calculated. According to literature [7], the elastic energy of original straight dislocation can be expressed as:

$$W_{str} = \frac{A_1 G b^2 (D+L)}{4\pi} \ln\left(\frac{R_o}{r_0}\right), \tag{S1}$$

Where G is the shear modulus, D is the precipitate diameter, L is the inter-precipitate spacing, $R_o$ is the integral range of the linear elasticity theory, $r_0$ is the radius of dislocation core, $A_1 = \cos^2\gamma + \frac{\sin^2\gamma}{1-\nu}$, $\gamma$ is the angle between the Burgers vector and the dislocation line, $\nu$ is Poisson's ratio. In addition, from Brown's work [8], the self-stress at discretized point position is $\tau_{self} = \frac{1}{b} \oint \frac{\sin\eta}{|r|^2} E dl$, where $b$ is magnitude of Burgers vector of the dislocation, **r** is the vector from discretized element d$l$ to this point, $\eta$ is the angle between **r** and the tangent at d$l$, E is the prelogarithmic factors in the elastic energy of an straight dislocation lying along **r**. Thus, the elastic energy for $i$th element can be expressed as: $dW_{ela} = \tau_{self}^i b R_i dl_i$, where $\tau_{self}^i$ is the self-stress at $i$th point, $R_i$ is the radius of the $i$th element, $dl_i$ is the length of $i$th discretized element. Naturally, the elastic energy for the whole bowing dislocation loop can be expressed as:

$$W_{ela} = \sum_{i=1}^{n} \tau_{self}^i b R_i dl, \tag{S2}$$

where n denotes the number of discretized elements.

In addition, according to the literature [9], the interaction energy between adjacent bowing loops can be expressed as:

$$W_{int}=\sum_{k=1}^{n}\frac{A_2 Gb^2(y_{k+1}-y_k)^2}{4\pi x_k}, \tag{S3}$$

where $(x_k, y_k)$ is the $k$th discretized point coordinate, $A_2=\frac{(1+v)\cos^2\gamma+(1-2v)\sin^2\gamma}{1-v}$.

Considering an area swept by the dislocation, work done by the external shear stress $\tau$ can be written as:

$$W_\tau=\frac{1}{2}\sum_{k=1}^{n}(y_{k+1}+y_k)(x_{k+1}-x_k)\tau b. \tag{S4}$$

Based on the above analysis, the dislocation system potential energy can be written as:

$$\Phi(\tau) = W_{ela} + W_{inti} + W_{inta} + W_\tau, \tag{S5}$$

$$=\sum_{i=1}^{n}\tau_{self}^i bR_i dl_i - \frac{A_1 Gb^2(D+L)}{4\pi}\ln\left(\frac{R_o}{r_0}\right) - \sum_{k=1}^{n}\frac{A_2 Gb^2(y_{k+1}-y_k)^2}{4\pi x_k} - \sum_{k=1}^{n/2}\frac{A_2 Gb^2(y_k)^2}{8\pi(D+L-2x_k)} - \frac{1}{2}\sum_{k=1}^{n}(y_{k+1}+y_k)(x_{k+1}-x_k)\tau b.$$

Furthermore, according to the self-stress is the product of bowing configuration and the unit $\frac{G}{4\pi(1-v)}$ [10, 11], as well as D and L in unit of $b$, thus, the system potential energy function can be transformation as:

$$\Phi(\tau)=Gb^3\left[\sum_{i=1}^{n}\frac{\tau_{self}^i}{G}R_i dl_i - \frac{A_1(D+L)}{4\pi}\ln\left(\frac{R_o}{r_0}\right) - \sum_{k=1}^{n}\frac{A_2(y_{k+1}-y_k)^2}{4\pi x_k} - \sum_{k=1}^{n/2}\frac{A_2(y_k)^2}{8\pi(D+L-2x_k)} - \frac{1}{2}\sum_{k=1}^{n}(y_{k+1}+y_k)(x_{k+1}-x_k)\tau/G\right]. \tag{S6}$$

Among them, the precipitates' geometric size (D and L), discretized point coordinate $(x_k, y_k)$, and the $\tau/G$ and $\tau_{self}^i/G$ are dimensionless.

**Part. SIII. The Derivation of Thermal Activation Energy**

According to a general activation energy expression, the activation energy as a function of the external shear stress $\tau$ can be written as:

$$\Delta\Phi(\tau) = gGb^3[1-(\tau/\tau_{BKS})^p]^q, \quad (S7)$$

where $gGb^3$ is a stress-free energy barrier, the exponents p and q are empirical parameters. $\Delta\Phi$ is used to characterize the energy barrier opposing dislocation motion.

Usually, when the energy barriers are of the order of electron-volt, the thermal energy favors the overcoming of this energy barriers. If the thermal energy meets the energy barriers, the dislocation can be thermally activated to by-pass the obstacles. The derivation of this thermal activation energy starts from the combination of Orowan equation and Arrhenius law based on the book by D. Caillard and J.-L. Martin [7]. According to Orowan equation, the plastic strain rate of crystals ($\dot{\varepsilon}$) is related to the mobile dislocation density ($\rho$), the average dislocation velocity ($\bar{v}$) and the Burger's vector ($b$) as:

$$\dot{\varepsilon} = \rho\bar{v}b. \quad (S8)$$

This expression of the plastic strain rate can be refined as follows. If $v_{id}$ is the vibration frequency of the dislocation segment, the probability for this dislocation segment to jump over the energy barrier can be expressed in the form of an Arrhenius law as:

$$P = v_{id}\exp(-\Delta\Phi/(k_BT)), \quad (S9)$$

where $k_B$ is the Boltzmann constant and T the absolute temperature. Among them, $v_{id}$ can be expressed considering that the wavelength of the vibration scales with $l$:

$$v_{id} = b\omega/l, \quad (S10)$$

where $\omega$ is the Debye frequency. And the average dislocation segment velocity ($\bar{v}$) is:

$$\bar{v} = P(A/l), \quad (S11)$$

where A is the area swept by dislocation segment with l between two successive obstacles. According to the above parameters, the plastic strain rate $\dot{\varepsilon}$ can be refined as:

$$\dot{\varepsilon} = \rho b P(A/l) = \rho b(A/l)(b\omega/l)\exp(-\Delta\Phi/(k_B T)) = \rho b\omega\lambda\exp(-\Delta\Phi/(k_B T)), \quad (S12)$$

Where $\lambda = Ab/l^2$ is the dislocations mean free path. Finally, we convert the above equation as:

$$\Delta\Phi = k_B T\ln(\rho b\lambda\omega/\dot{\varepsilon}). \quad (S13)$$

Taking Eq. S13 as the thermal activation energy, its need to meet the energy barriers in the value to make the dislocation jump over obstacles. Therefore, combining Eq. S7 with Eq. S13, we get the expression of external shear stress $\tau$ as a function of the plastic strain rate $\dot{\varepsilon}$, mobile dislocation density $\rho$, Debye frequency $\omega$ and temperature T as:

$$gGb^3[1-(\tau/\tau_{BKS})^p]^q = k_B T\ln(\rho b\lambda\omega/\dot{\varepsilon}). \quad (S14)$$

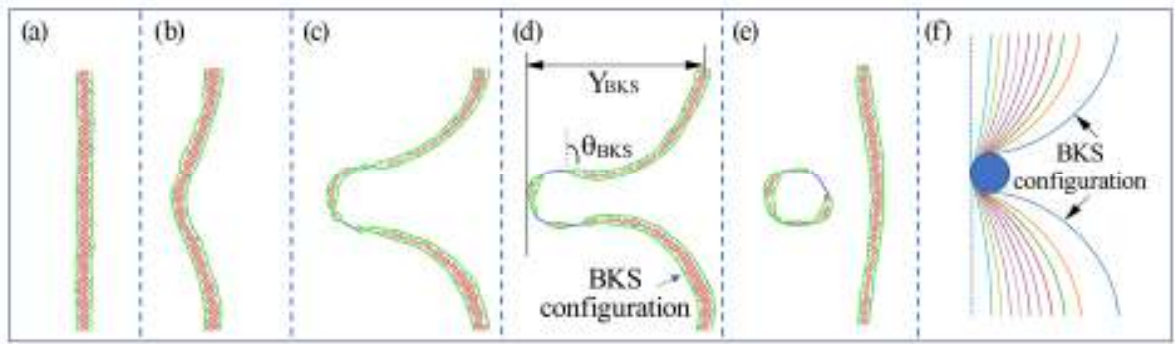

**Fig. S1. Comparison between MD simulated snapshots and this work numerical calculated bowing pathway.** (a) The snapshot of the dislocation configuration when the impenetrable precipitate is far away the dislocation. In this simulation, the precipitate diameter is D = 2.56 nm, inter-precipitate spacing L = 25.6 nm. (b) The snapshot of the dislocation configuration when the dislocation is pinned by the precipitate. (c) The snapshot of equilibrium configuration. (d) The snapshot of BKS configuration corresponding BKS stress. The configuration is characterized by tangent angle $\theta_{BKS}$ and bow-out height $Y_{BKS}$. (e) The snapshot of forming Orowan loop. (f) The bowing pathway calculated by this work numerical method.

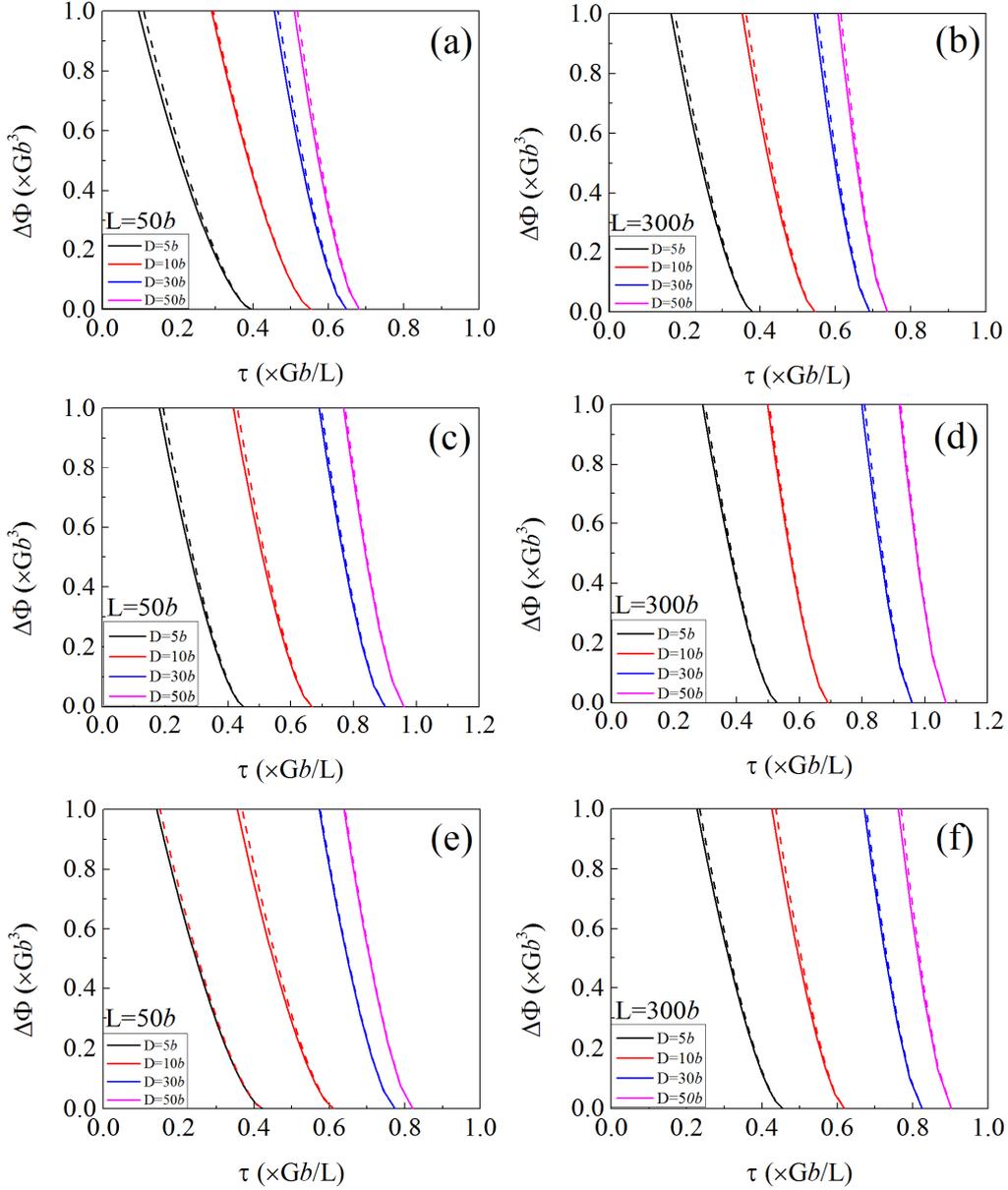

**Fig. S2. Comparison of activation energy from numerical method and analytical method.** (a) The comparison of activation energy from numerical method and analytical method for edge dislocation in L = 50$b$ case. (b) The comparison of activation energy from numerical method and analytical method for edge dislocation in L = 300$b$ case. (c) The comparison of activation energy from numerical method and analytical method for screw dislocation in L = 50$b$ case. (d) The comparison of activation energy from numerical method and analytical method for screw dislocation in L = 300$b$ case. (e) The comparison of activation energy from numerical method and analytical method for mixed 45° dislocation in L = 50$b$ case. (f) The comparison of activation energy from numerical method and analytical method for mixed 45° dislocation in L = 300$b$ case. The solid curves are calculated by the numerical method, and the dashed curves are calculated by the analytical method.

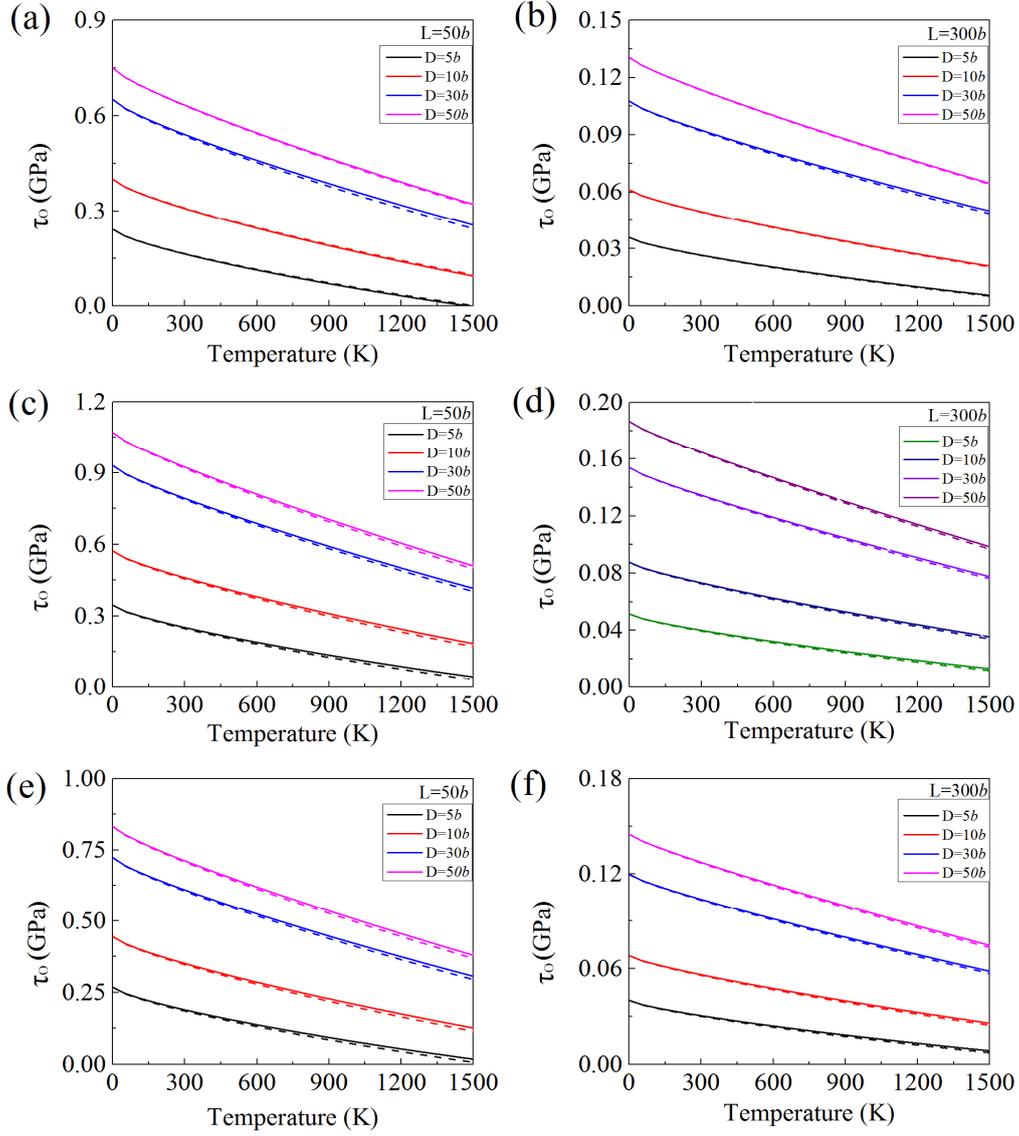

**Fig. S3. Comparison of thermal-activated Orowan stress $\tau_o$ deduced from numerical method and analytical method.** (a) The comparison of $\tau_o$ deduced from numerical method and analytical method for edge dislocation in L = 50*b* case. (b) The comparison of $\tau_o$ deduced from numerical method and analytical method for edge dislocation in L = 300*b* case. (c) The comparison of $\tau_o$ deduced from numerical method and analytical method for screw dislocation in L = 50*b* case. (d) The comparison of $\tau_o$ deduced from numerical method and analytical method for screw dislocation in L = 300*b* case. (e) The comparison of $\tau_o$ deduced from numerical method and analytical method for mixed 45° dislocation in L = 50*b* case. (f) The comparison of $\tau_o$ deduced from numerical method and analytical method for mixed 45° dislocation in L = 300*b* case. The solid curves are deduced from numerical method, and the dashed curves are deduced from analytical method.

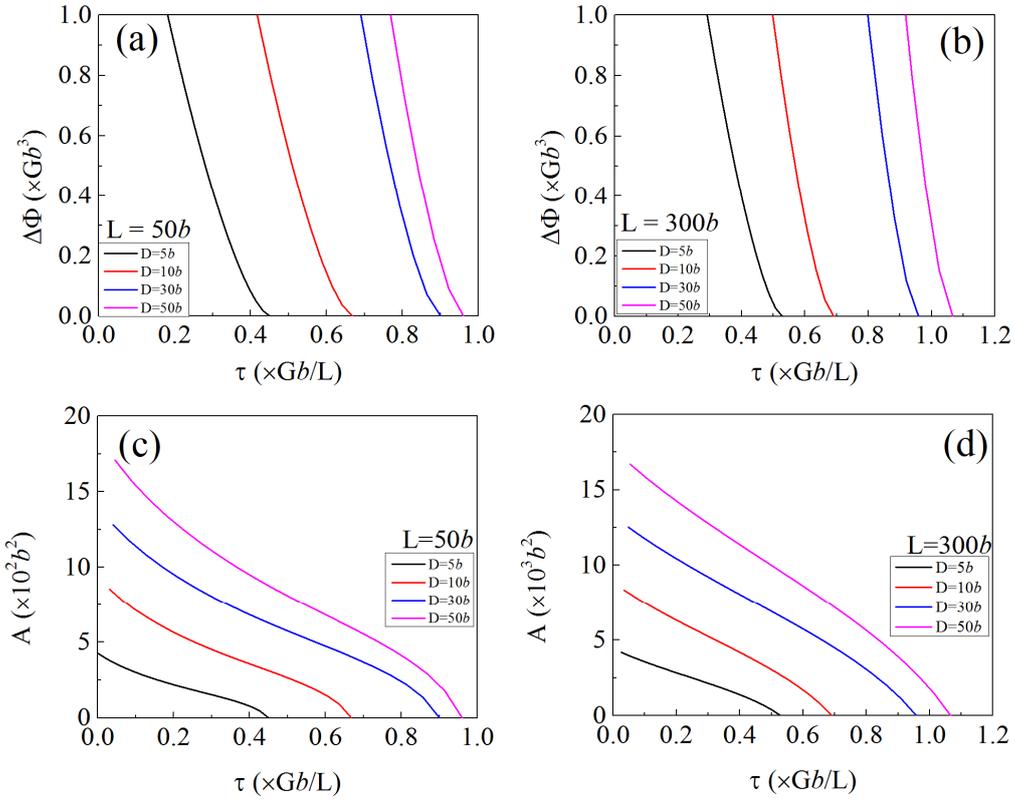

**Fig. S4. Effect of obstacle scale on activation energy and activation area for screw dislocation.** (a) Effect of precipitate diameter on the activation energy for small obstacle scale (L = 50$b$) case. (b) Effect of precipitate diameter on the activation energy for large obstacle scale (L = 300$b$) case. (c) Effect of precipitate diameter on activation area for small obstacle scale (L = 50$b$) case. (d) Effect of precipitate diameter on activation area for large obstacle scale (L = 300$b$) case.

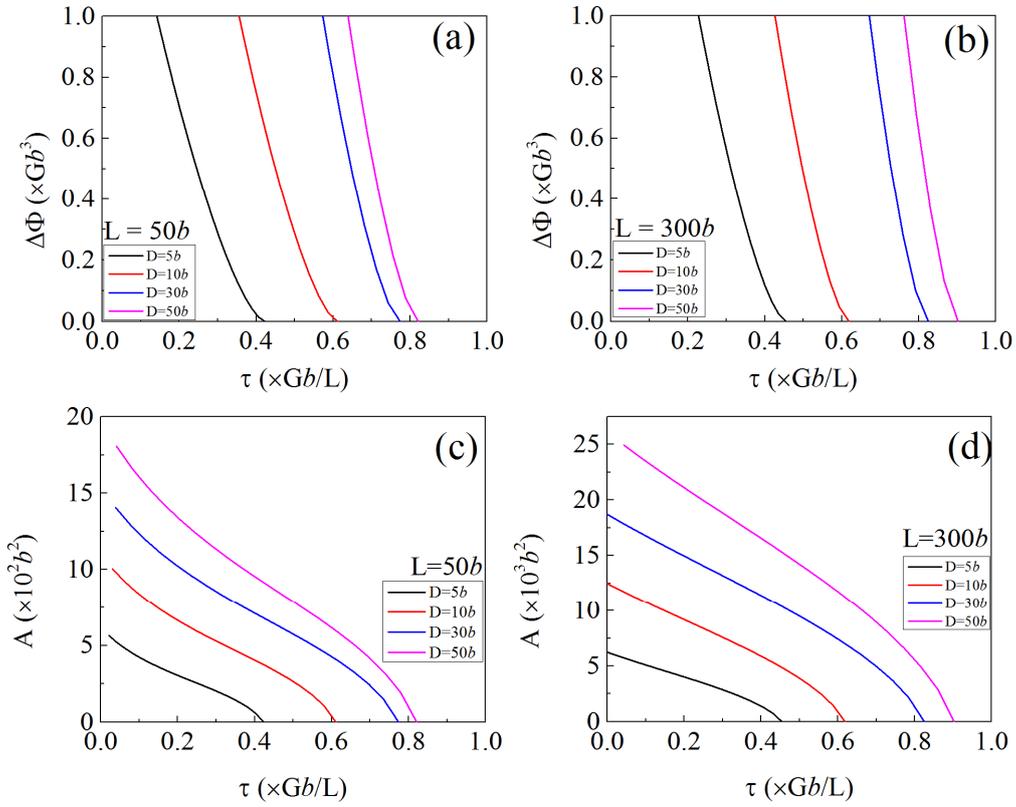

**Fig. S5. Effect of obstacle scale on activation energy and activation area for mix 45° dislocation.** (a) Effect of precipitate diameter on the activation energy for small obstacle scale (L = 50$b$) case. (b) Effect of precipitate diameter on the activation energy for large obstacle scale (L = 300$b$) case. (c) Effect of precipitate diameter on activation area for small obstacle scale (L = 50$b$) case. (d) Effect of precipitate diameter on activation area for large obstacle scale (L = 300$b$) case.

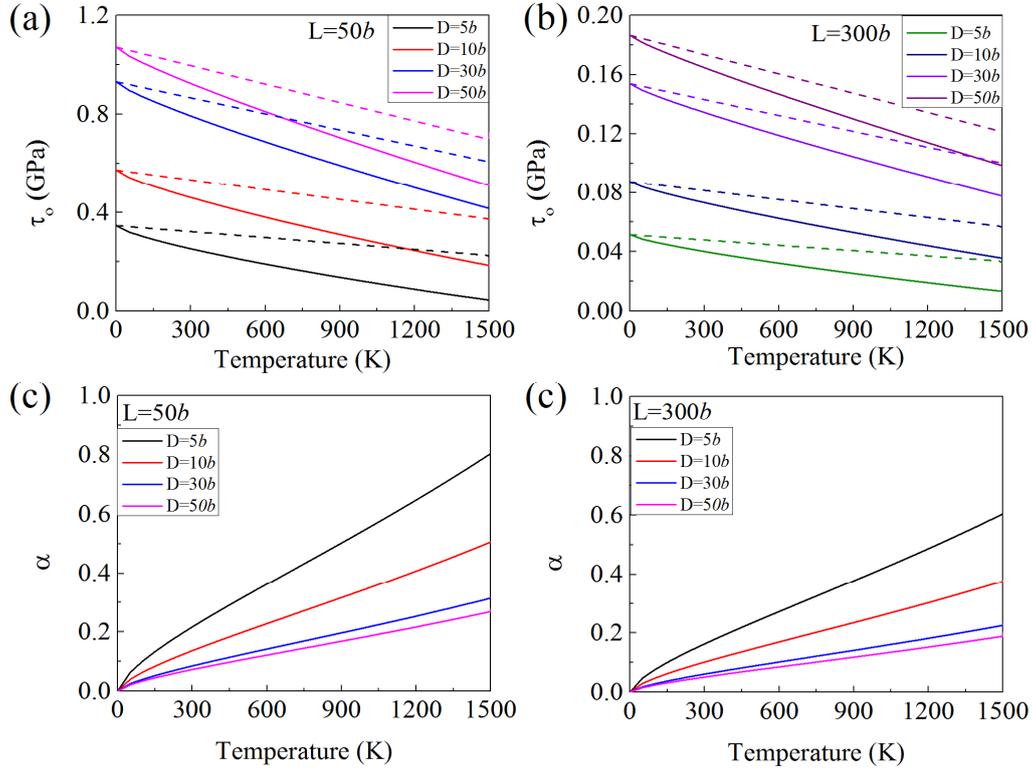

**Fig. S6. Effect of obstacle scale on thermal-activated Orowan stress $\tau_o$ and thermal activation factor α for screw dislocation.** (a) Effect of precipitates diameter on $\tau_o$ for small obstacle scale (L = 50*b*) case. In this calculation, the mobile dislocations density is $\rho = 10^{12}$ m$^{-2}$, the strain rate is $\dot{\varepsilon} = 10^{-2}$ s$^{-1}$, shear module as a function of temperature T is G = 75[1-0.5(T-300)/1700] GPa, dislocation mean free path is $\lambda = 10b$, Debye frequency is $\omega = 10^{13}$ THz. The solid curves are thermal-activated Orowan stress $\tau_o$ calculated by our model, while the dashed lines are Orowan stress without consideration of thermal activation, $\beta\tau_{BKS}$. (b) Effect of precipitates diameter on $\tau_o$ for large obstacle scale (L = 300*b*) case. (c) Effect of precipitate diameter on α for small obstacle scale (L = 50*b*) case. (d) Effect of precipitates diameter on α for large obstacle scale (L = 300*b*) case.

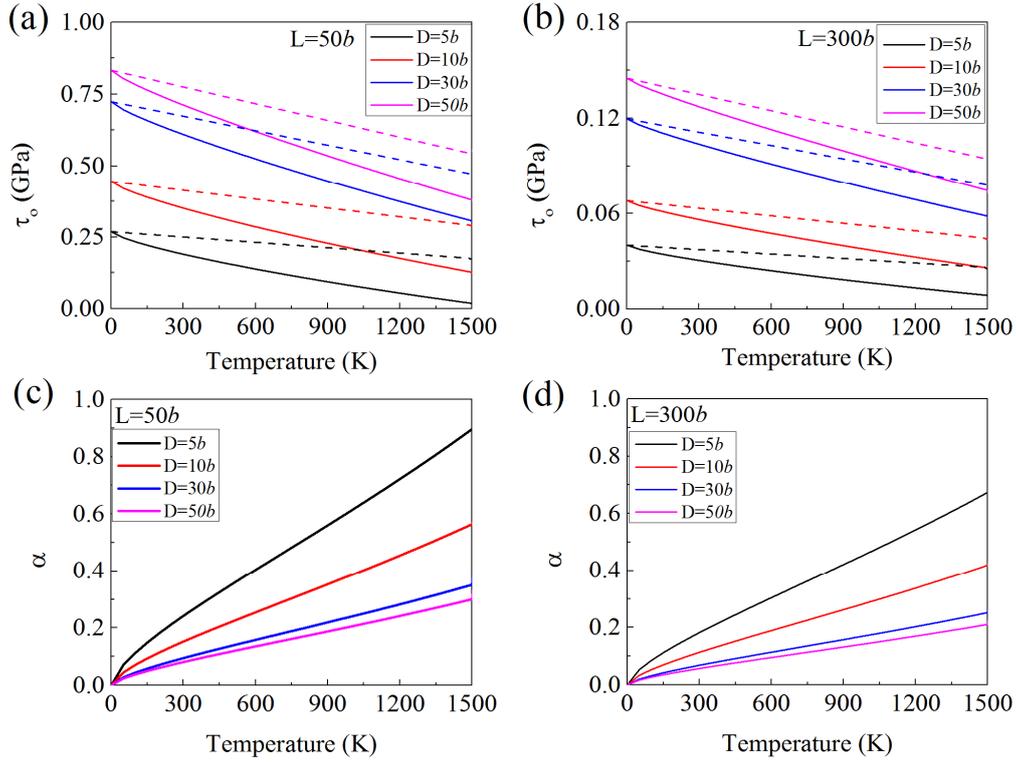

**Fig. S7. Effect of obstacle scale on thermal-activated Orowan stress $\tau_o$ and thermal activation factor α for mix 45° dislocation.** (a) Effect of precipitates diameter on $\tau_o$ for small obstacle scale (L = 50$b$) case. In this calculation, the mobile dislocations density is $\rho = 10^{12}$ m$^{-2}$, the strain rate is $\dot{\varepsilon} = 10^{-2}$ s$^{-1}$, shear module as a function of temperature T is G = 75[1-0.5(T-300)/1700] GPa, dislocation mean free path is $\lambda = 10b$, Debye frequency is $\omega = 10^{13}$ THz. The solid curves are thermal-activated Orowan stress $\tau_o$ calculated by our model, while the dashed lines are Orowan stress without consideration of thermal activation, $\beta\tau_{BKS}$ [11]. (b) Effect of precipitates diameter on $\tau_o$ for large obstacle scale (L = 300$b$) case. (c) Effect of precipitate diameter on α for small obstacle scale (L = 50$b$) case. (d) Effect of precipitates diameter on α for large obstacle scale (L = 300$b$) case.

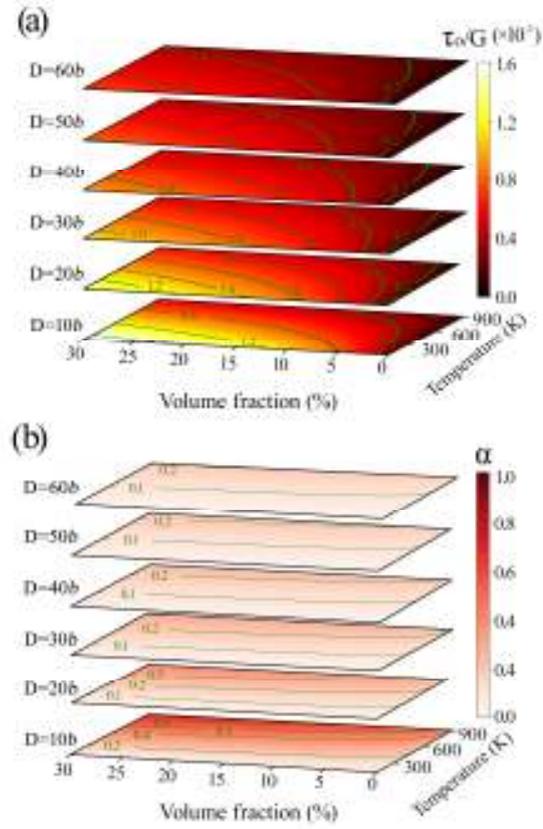

**Fig. S8. Thermal-activated Orowan stress $\tau_o$ and thermal activation factor α under variable precipitate volume fraction for screw dislocation.** (a) Normalized thermal-activated Orowan stress $\tau_o/G$ as a function of precipitate diameter, precipitate volume fraction, and temperature. In this calculation, the mobile dislocations density is $\rho = 10^{12}$ m$^{-2}$, the strain rate is $\dot{\varepsilon} = 10^{-2}$ s$^{-1}$, shear modulus as a function of temperature T is $G = 75[1-0.5(T-300)/1700]$ GPa, dislocation mean free path is $\lambda = 10b$, Debye frequency is $\omega = 10^{13}$ THz. (b) α as a function of precipitate diameter, precipitate volume fraction, and temperature.

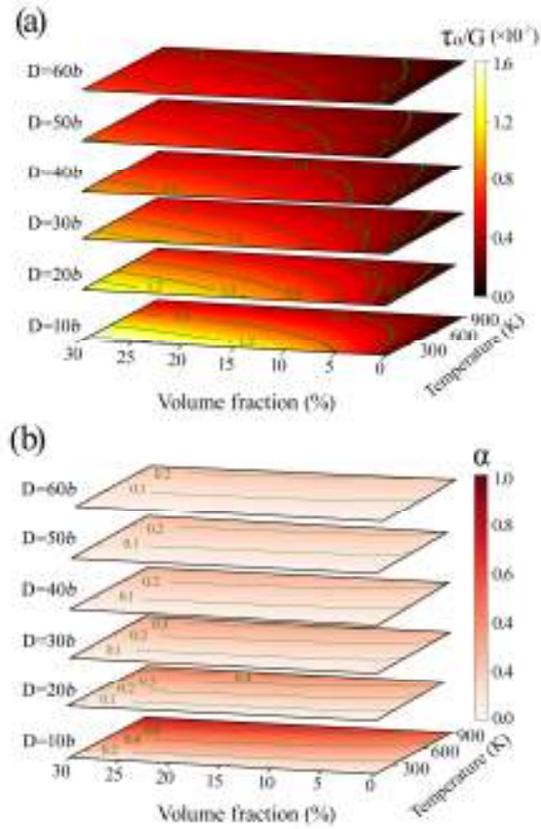

**Fig. S9. Thermal-activated Orowan stress $\tau_o$ and thermal activation factor α under variable precipitate volume fraction for mix 45° dislocation.** (a) Normalized thermal-activated Orowan stress $\tau_o/G$ as a function of precipitate diameter, precipitate volume fraction, and temperature. In this calculation, the mobile dislocations density is $\rho = 10^{12}$ m$^{-2}$, the strain rate is $\dot{\varepsilon} = 10^{-2}$ s$^{-1}$, shear modulus as a function of temperature T is $G = 75[1-0.5(T-300)/1700]$ GPa, dislocation mean free path is $\lambda = 10b$, Debye frequency is $\omega = 10^{13}$ THz. (b) α as a function of precipitate diameter, precipitate volume fraction, and temperature.

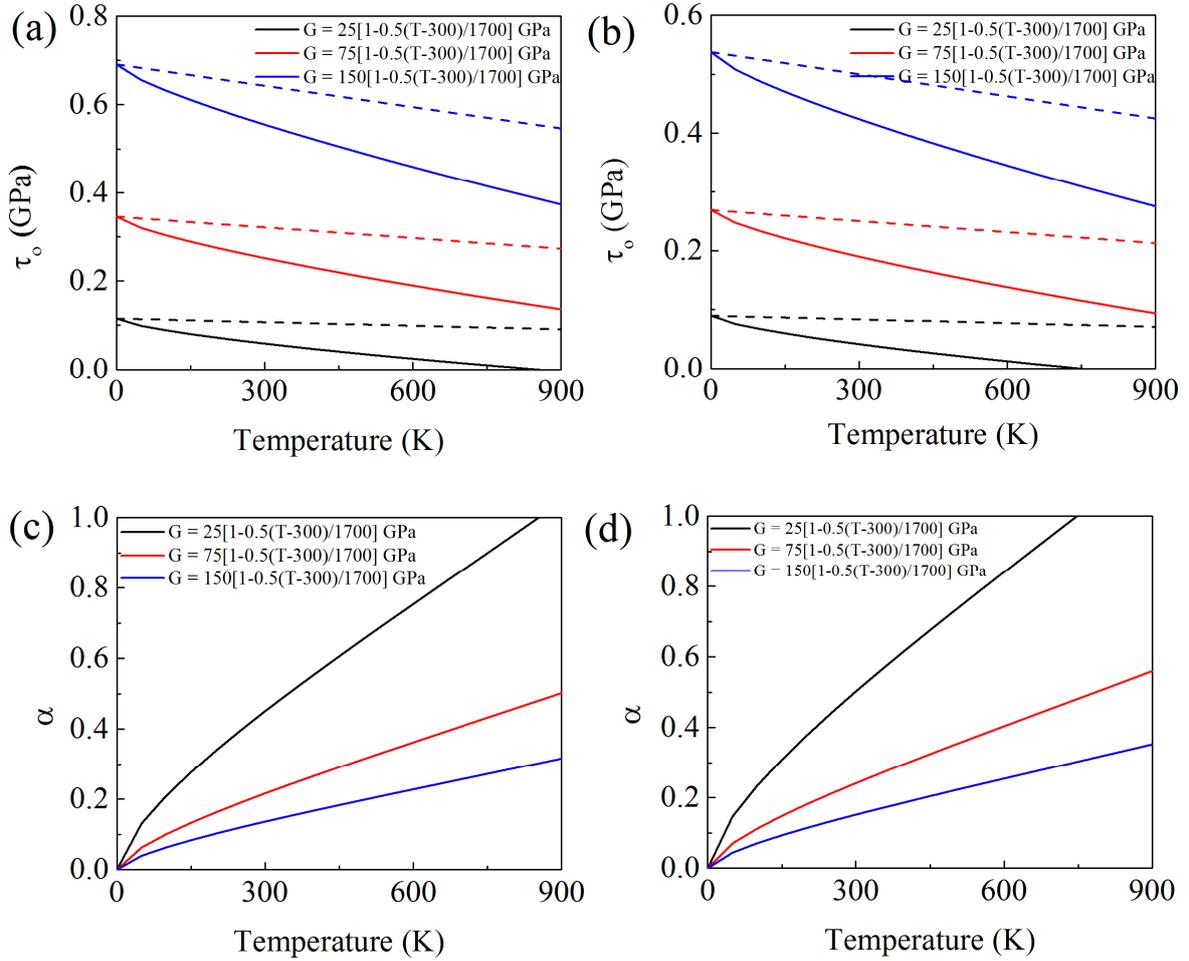

**Fig. S10. Thermal-activated Orowan stress τ₀ and thermal activation factor α under variable shear modulus.** (a) Variations in $\tau_o$ with temperature under different shear modulus cases for screw dislocation. In this calculation, the mobile dislocations density is $\rho = 10^{12}$ m$^{-2}$, and the strain rate is $\dot{\varepsilon} = 10^{-2}$ s$^{-1}$, precipitate diameter is D = 5$b$, inter-precipitate spacing is L = 50$b$, dislocation mean free path is $\lambda = 10b$, Debye frequency is $\omega = 10^{13}$ THz. The solid curves are thermal-activated Orowan stress $\tau_o$ calculated by our model, while the dashed lines are Orowan stress without consideration of thermal activation, $\beta\tau_{BKS}$. (b) Variations in $\tau_o$ with temperature under different shear modulus cases for mix 45° dislocation. In this calculation, the mobile dislocations density is $\rho = 10^{12}$ m$^{-2}$, and the strain rate is $\dot{\varepsilon} = 10^{-2}$ s$^{-1}$, precipitate diameter is D = 5$b$, inter-precipitate spacing is L = 50$b$, dislocation mean free path is $\lambda = 10b$, Debye frequency is $\omega = 10^{13}$ THz. The solid curves are thermal-activated Orowan stress $\tau_o$ calculated by our model, while the dashed lines are Orowan stress without consideration of thermal activation, $\beta\tau_{BKS}$. (c) Variations in α with temperature under different shear modulus cases for screw dislocation. (d) Variations in α with temperature under different shear modulus cases for mix 45° dislocation.

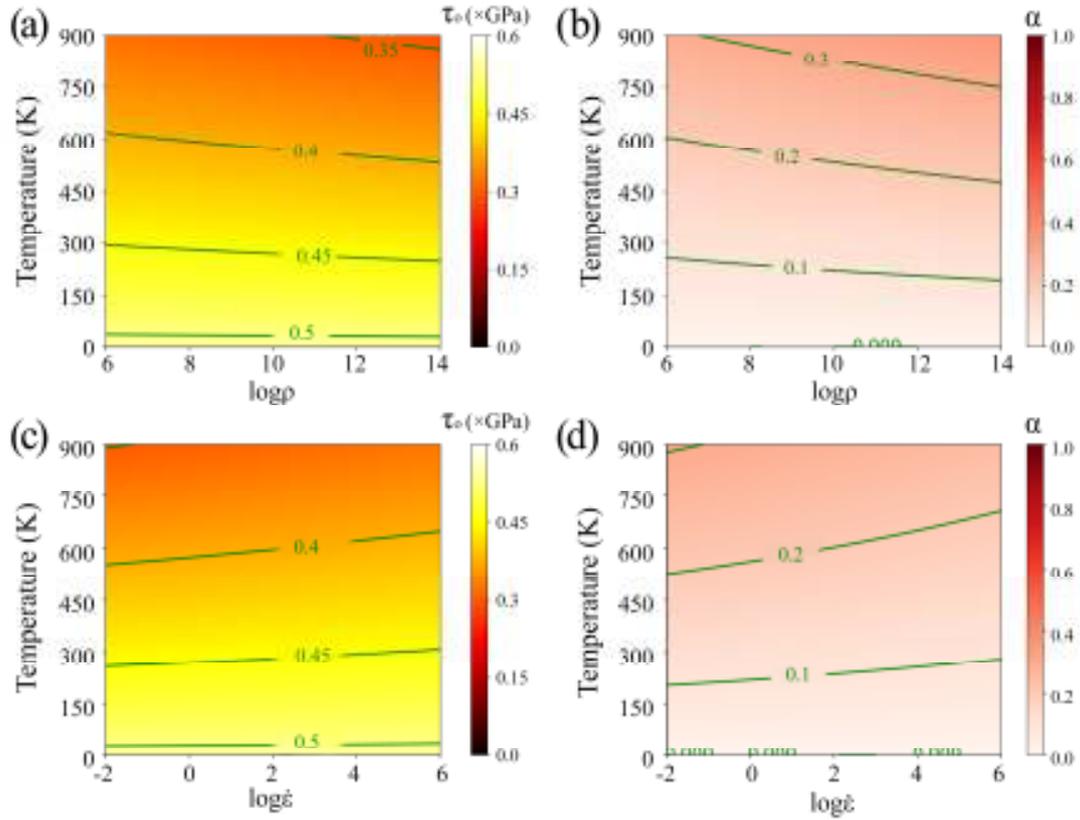

**Fig. S11. Thermal-activated Orowan stress $\tau_o$ and thermal activation factor $\alpha$ as a function of mobile dislocations density and strain rate for screw dislocation.** (a) $\tau_o$ as a function of dislocation density and temperature. In this calculation, mobile dislocations density is varying from $10^6$ to $10^{14}$ m$^{-2}$, the temperature is varying from 0 to 900 K, and stain rate is $\dot{\varepsilon} = 10^{-2}$ s$^{-1}$, precipitate diameter is D = 10$b$, inter-precipitate spacing is L = 50$b$, shear modulus as a function of temperature T is G = 75[1-0.5(T-300)/1700] GPa, dislocation mean free path is $\lambda = 10b$, Debye frequency is $\omega = 10^{13}$ THz. (b) $\alpha$ as a function of mobile dislocations density and temperature. (c) $\tau_o$ as a function of strain rate and temperature. In this calculation, the strain rate is varying from $10^{-2}$ to $10^6$ s$^{-1}$, and the temperature is varying from 0 to 900 K, mobile dislocations density is $\rho = 10^{12}$ m$^{-2}$, precipitate diameter is D = 10$b$, inter-precipitate spacing is L = 50$b$, the shear modulus as a function of temperature T is G = 75[1-0.5(T-300)/1700] GPa, dislocation mean free path is $\lambda = 10b$, Debye frequency is $\omega = 10^{13}$ THz. (d) $\alpha$ as a function of strain rate and temperature.

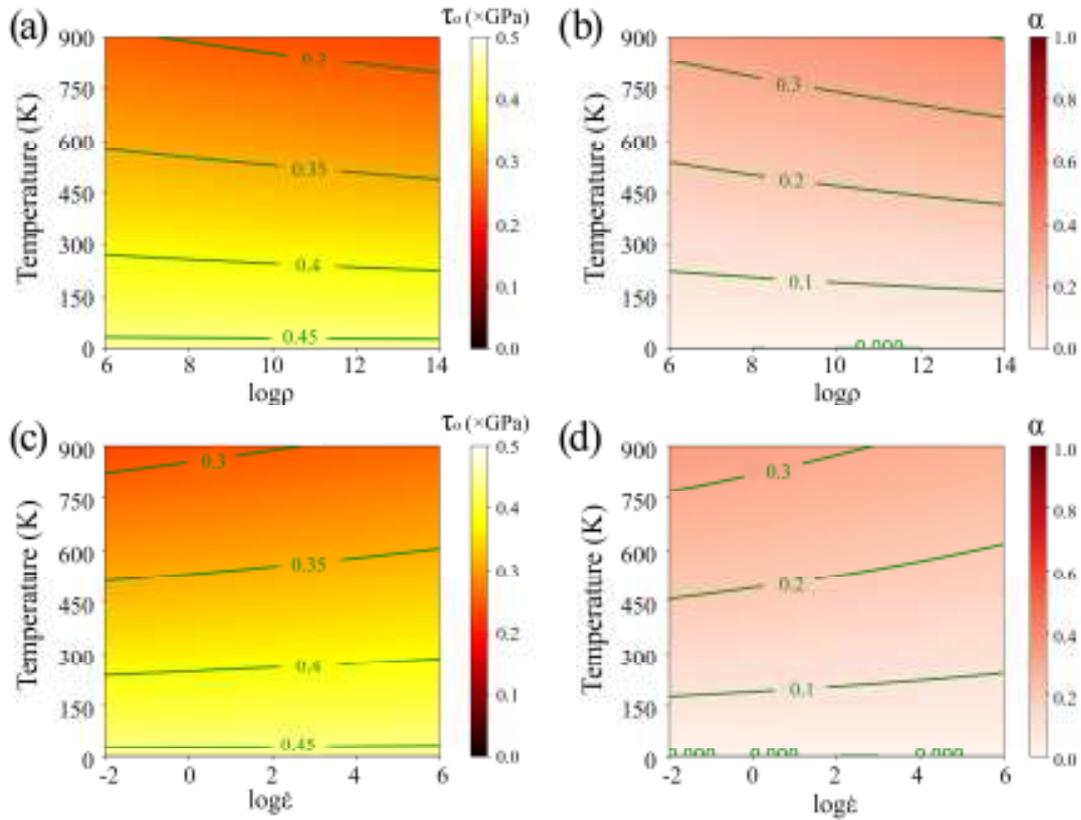

**Fig. S12. Thermal-activated Orowan stress $\tau_o$ and thermal activation factor $\alpha$ as a function of mobile dislocations density and strain rate for mix 45° dislocation.** (a) $\tau_o$ as a function of dislocation density and temperature. In this calculation, mobile dislocations density is varying from $10^6$ to $10^{14}$ m$^{-2}$, the temperature is varying from 0 to 900 K, and stain rate is $\dot{\varepsilon} = 10^{-2}$ s$^{-1}$, precipitate diameter is D = 10$b$, inter-precipitate spacing is L = 50$b$, shear modulus as a function of temperature T is G = 75[1-0.5(T-300)/1700] GPa, dislocation mean free path is $\lambda$ = 10$b$, Debye frequency is $\omega$ = $10^{13}$ THz. (b) $\alpha$ as a function of mobile dislocations density and temperature. (c) $\tau_o$ as a function of strain rate and temperature. In this calculation, the strain rate is varying from $10^{-2}$ to $10^6$ s$^{-1}$, and the temperature is varying from 0 to 900 K, mobile dislocations density is $\rho$ = $10^{12}$ m$^{-2}$, precipitate diameter is D = 10$b$, inter-precipitate spacing is L = 50$b$, the shear modulus as a function of temperature T is G = 75[1-0.5(T-300)/1700] GPa, dislocation mean free path is $\lambda$ = 10$b$, Debye frequency is $\omega$ = $10^{13}$ THz. (d) $\alpha$ as a function of strain rate and temperature.

**Table S1. Bowing pathway parameters of edge, screw and mixed 45° dislocations under different precipitates' geometric size.** $A_{BKS}$, $\theta_{BKS}$, $Y_{BKS}$, $P_{BKS}$ is swept area, tangent angle and height and perimeter of the bowing loop for BKS configuration. $A_{BKS}$ in units of $L^2$, $Y_{BKS}$ and $P_{BKS}$ in units of L.

| L | D Type Item | 5b Edge | 5b Screw | 5b Mix | 10b Edge | 10b Screw | 10b Mix | 30b Edge | 30b Screw | 30b Mix | 50b Edge | 50b Screw | 50b Mix | 100b Edge | 100b Screw | 100b Mix |
|---|---|---|---|---|---|---|---|---|---|---|---|---|---|---|---|---|
| 50b | $\theta_{BKS}$ | 86.3 | 90.7 | 96.2 | 92.6 | 96.3 | 109 | 104.6 | 97.8 | 107.6 | 96.4 | 98.9 | 97.9 | 100.5 | 106.1 | 106 |
| 50b | $A_{BKS}$ | 0.33 | 0.18 | 0.21 | 0.57 | 0.34 | 0.33 | 1.19 | 0.70 | 0.41 | 0.92 | 0.82 | 0.63 | 1.13 | 0.91 | 1.06 |
| 50b | $P_{BKS}$ | 1.31 | 1.12 | 1.14 | 1.62 | 1.23 | 1.01 | 1.48 | 1.39 | 1.24 | 1.62 | 1.33 | 1.33 | 1.80 | 1.28 | 1.42 |
| 50b | $Y_{BKS}$ | 0.42 | 0.23 | 0.26 | 0.69 | 0.41 | 0.33 | 0.88 | 0.77 | 0.48 | 1.06 | 0.89 | 0.72 | 1.13 | 0.93 | 0.99 |
| 100b | $\theta_{BKS}$ | 89.8 | 92.9 | 95.4 | 90.0 | 91.7 | 107 | 90.0 | 92.7 | 120 | 92.3 | 93.8 | 99.9 | 92.5 | 91.1 | 99.7 |
| 100b | $A_{BKS}$ | 0.2 | 0.13 | 0.19 | 0.44 | 0.25 | 0.30 | 0.56 | 0.44 | 0.378 | 0.63 | 0.56 | 0.60 | 0.89 | 0.79 | 0.76 |
| 100b | $P_{BKS}$ | 1.25 | 1.07 | 1.14 | 1.52 | 1.14 | 1.10 | 1.32 | 1.20 | 1.24 | 1.57 | 1.28 | 1.18 | 1.61 | 1.28 | 1.32 |
| 100b | $Y_{BKS}$ | 0.34 | 0.16 | 0.27 | 0.56 | 0.29 | 0.44 | 0.70 | 0.51 | 0.479 | 0.77 | 0.63 | 0.62 | 1.03 | 0.87 | 0.89 |
| 300b | $\theta_{BKS}$ | 93.6 | 90.0 | 97.3 | 95.9 | 94.7 | 115 | 93.2 | 91.9 | 107 | 90.8 | 94.7 | 99.5 | 90.4 | 94.9 | 94.7 |
| 300b | $A_{BKS}$ | 0.19 | 0.09 | 0.10 | 0.29 | 0.14 | 0.15 | 0.50 | 0.24 | 0.16 | 0.46 | 0.33 | 0.32 | 0.57 | 0.45 | 0.37 |
| 300b | $P_{BKS}$ | 1.16 | 1.04 | 1.08 | 1.34 | 1.08 | 1.06 | 1.24 | 1.07 | 1.11 | 1.53 | 1.23 | 1.14 | 1.56 | 1.28 | 1.30 |
| 300b | $Y_{BKS}$ | 0.26 | 0.12 | 0.15 | 0.39 | 0.18 | 0.15 | 0.63 | 0.29 | 0.34 | 0.59 | 0.39 | 0.31 | 0.71 | 0.52 | 0.56 |